\documentclass[acmsmall,nonacm]{acmart}

\renewcommand\footnotetextcopyrightpermission[1]{}
\acmJournal{PACMPL}
\acmVolume{1}
\acmNumber{OOPSLA} 
\acmArticle{1}
\acmYear{2022}
\acmMonth{10}
\startPage{1}

\setcopyright{none}

\usepackage{pifont}
\usepackage{paralist}
\usepackage{booktabs}
\usepackage{threeparttable}
\usepackage{enumitem}    
\usepackage[greek,english]{babel}
\usepackage{ifthen}
\usepackage{xspace}
\usepackage{fancybox}
\usepackage{marginnote}

\usepackage[many]{tcolorbox}
\usepackage{multirow}
\usepackage{mathtools}
\usepackage{datetime}
\usepackage{algorithm}
\usepackage[normalem]{ulem}
\usepackage[nomessages]{fp}
\usepackage[noend]{algpseudocode}
\usepackage{graphicx}
\usepackage{caption}
\usepackage{balance}
\usepackage{makecell}
\usepackage{wrapfig}
\usepackage{listings}
\usepackage{xcolor}
\usepackage{expl3, xparse, siunitx}
\usepackage{alltt}
\ExplSyntaxOn
\usepackage{cleveref}
\usepackage{subcaption}
\usepackage{tikz}
\usepackage{circledsteps}
\usepackage{soul}
\AtEndPreamble{%
	\theoremstyle{acmdefinition}
	\newtheorem{remark}[theorem]{Remark}}

\NewDocumentCommand { \calcnum } { O{} m }
  { \num [  round-mode=places , round-precision=2 , group-separator={,}, group-minimum-digits=4, #1] { \fp_to_decimal:n {#2} } }
\ExplSyntaxOff

\newboolean{showcomments} 
\setboolean{showcomments}{true}
\ifthenelse{\boolean{showcomments}}
{\newcommand{\nb}[2]{
		\fbox{\bfseries\sffamily\scriptsize#1}
		{\sf\small$\blacktriangleright$\textit{\textcolor{blue}{#2}}$\blacktriangleleft$}
	}
	
}
{\newcommand{\nb}[2]{}}

\newcommand\ameya[1]{\nb{Ameya}{#1}}

\newcommand\mohammad[1]{\nb{Mohammad}{#1}}
\newcommand\yasharth[1]{\nb{Yasharth}{#1}}
\newcommand\priyanshu[1]{\nb{\textcolor{teal} {Priyanshu}}{#1}}

\newcommand{\todo}[1]{\textcolor{magenta}{\nb{TODO:}{#1}}}

\newcommand{\ignore}[1]{}

\newcommand{\hide}[1]{}

\newcommand{\code}[1]{\text{\texttt{\fontsize{9.5}{10}\selectfont #1}}}
\newcommand{\smcode}[1]{\text{\texttt{\fontsize{7.5}{8}\selectfont #1}}}

\definecolor{findingsbox-bg-color}{gray}{0.90}
\newtcbox{\findingsbox}{colback=findingsbox-bg-color, boxrule=0.2pt, arc=2pt, boxsep=0pt, left=5pt, right=5pt, top=5pt, bottom=5pt}

\cornersize{.15}

\crefname{lstlisting}{listing}{listings}

\definecolor{dkgreen}{rgb}{0,0.6,0}
\definecolor{gray}{rgb}{0.5,0.5,0.5}
\definecolor{mauve}{rgb}{0.58,0,0.82}

\newcommand{\technique}{\textsc{Overwatch}\xspace}
\newcommand{\noOfsessions}{682\xspace}
\newcommand{\noOfDocuments}{425\xspace}
\newcommand{\totalVersions}{134,545\xspace}

\newcommand{\numberOfDevelopers}{12\xspace}

\newcommand{\noOfInsertSnippets}{10\xspace}
\newcommand{\noOfInsertSnippetsPatterns}{15\xspace}

\newcommand{\arsays}[1]{{\color{pink} {\textbf{AR:} #1}}}

\newcommand{\edit}{\mathsf{ed}}
\newcommand{\editSeq}{\mathsf{edSeq}}
\newcommand{\editSequences}{\mathsf{edSeqs}}

\newcommand{\Edits}{\mathsf{Edits}}
\newcommand{\edits}{\mathsf{edits}}

\newcommand{\version}{\mathsf{v}}

\newcommand{\vpre}{\version_\mathsf{pre}}
\newcommand{\vpost}{\version_\mathsf{post}}
\newcommand{\tpltpre}{\tplt_\mathsf{pre}}
\newcommand{\tpltpost}{\tplt_\mathsf{post}}

\newcommand{\InsertChild}{\mathsf{Insert}}
\newcommand{\DeleteChild}{\mathsf{Delete}}
\newcommand{\Update}{\mathsf{Update}}

\newcommand{\parent}{\mathsf{parent}}
\newcommand{\child}{\mathsf{child}}
\newcommand{\old}{\mathsf{old}}
\newcommand{\new}{\mathsf{new}}

\newcommand{\addCode}[1]{{\color{green!50!black}{#1}}}
\newcommand{\delCode}[1]{{\color{red!50!black}{#1}}}
\newcommand{\minus}[1]{\delCode{- #1}}
\newcommand{\plus}[1]{\addCode{+ #1}}

\newcommand{\Session}{\mathsf{Trace}}
\newcommand{\Sessions}{\mathsf{Traces}}

\newcommand{\seq}{\mathsf{seq}}

\newcommand{\ASTnode}{\mathsf{node}}

\newcommand{\graphnode}{v}
\newcommand{\edseqnumber}{s}

\newcommand{\Partitions}{\mathsf{Partitions}}
\newcommand{\Partition}{\mathsf{P}}

\newcommand{\lstbg}[3][0pt]{{\fboxsep#1
  \colorbox{#2}{\strut \ensuremath{#3}}}}

\lstdefinelanguage{diff}{
  basicstyle=\ttfamily\footnotesize,
  morecomment=[f][\lstbg{red!20}]-,
  morecomment=[f][\lstbg{green!20}]+,
  morecomment=[f][\textit]{@@},
  morecomment=[f][\textbf]{//},
}

\newcommand{\hole}[0]{\mathsf{H}}
\newcommand{\tplt}[0]{\mathsf{t}}
\newcommand{\etplt}[0]{\mathsf{et}}
\newcommand{\tpltseq}[0]{\mathsf{TS}}
\newcommand{\tmprel}[0]{\mathsf{Preds}}

\newcommand{\localize}[0]{\mathsf{Localize}}
\newcommand{\substitution}[0]{\sigma}

\newcommand{\dependency}{\mathsf{F}}
\newcommand{\pre}{\mathsf{pre}}
\newcommand{\post}{\mathsf{post}}

\newcommand{\sketch}{\mathsf{sk}}
\newcommand{\spec}{\mathsf{spec}}
\newcommand{\paththreshold}{2}

\newcommand{\mytodoviolet}[1]{\textcolor{violet}{\ding{46}~{\sf}~#1}}

\newcommand{\yh}[1]{\nb{\mytodoviolet{yuhao}}{#1}}

\usetikzlibrary{calc,fit,positioning,shapes.misc,shapes.geometric,shapes,arrows,chains,mindmap,trees,backgrounds}

\usepackage{threeparttablex}
\usepackage{longtable}
\usepackage{pdflscape}

\begin{document}

\title{Overwatch: Learning Patterns in Code Edit Sequences}

\author{Yuhao Zhang}
\email{yuhaoz@cs.wisc.edu}
\authornote{This work was done when these authors were 
employed at Microsoft}
\authornote{Equal contribution}
\affiliation{%
  \institution{University of Wisconsin-Madison}
  \country{USA}
}

\author{Yasharth Bajpai}
\email{t-yabajpai@microsoft.com}
\authornotemark[2]
\affiliation{%
  \institution{Microsoft}
  \country{India}
}

\author{Priyanshu Gupta}
\email{priyansgupta@microsoft.com}
\authornotemark[2]
\affiliation{%
  \institution{Microsoft}
  \country{India}
}

\author{Ameya Ketkar}
\authornotemark[1]
\authornotemark[2]
\affiliation{%
  \institution{Uber}
  \country{USA}
}
\email{ketkara@uber.com}

\author{Miltiadis Allamanis}
\affiliation{%
  \institution{Microsoft Research}
  \country{UK}
}
\email{miltos@allamanis.com}

\author{Titus Barik}
\affiliation{%
  \institution{Apple}
  \country{USA}
}
\authornotemark[1]
\email{tbarik@acm.org}

\author{Sumit Gulwani}
\affiliation{%
  \institution{Microsoft}
  \country{USA}
}
\email{sumitg@microsoft.com}

\author{Arjun Radhakrishna}
\affiliation{%
  \institution{Microsoft}
  \country{USA}
}
\email{arradha@microsoft.com}

\author{Mohammad Raza}
\affiliation{%
  \institution{Microsoft}
  \country{USA}
}
\email{moraza@microsoft.com}

\author{Gustavo Soares}
\affiliation{%
  \institution{Microsoft}
  \country{USA}
}
\email{gsoares@microsoft.com}

\author{Ashish Tiwari}
\affiliation{%
  \institution{Microsoft}
  \country{USA}
}
\email{astiwar@microsoft.com}

\begin{abstract}
Integrated Development Environments (IDEs) provide tool support
to automate many source code editing tasks. 
Traditionally, IDEs use only the spatial context, i.e., the location where
the developer is editing, to generate candidate edit recommendations.
However, spatial context alone is often not sufficient to confidently
predict the developer's next edit, and thus IDEs generate many
suggestions at a location.
Therefore, IDEs generally do not actively offer suggestions and instead, the
developer is usually required to click on a specific icon or
menu and then select from a large list of
potential suggestions.
As a consequence, developers often miss the opportunity to use the tool support because they are not aware it exists or forget to use it.  
%
%

To better understand common patterns in developer behavior
and produce better edit recommendations,
we can additionally use the \emph{temporal context},
i.e., the edits that a developer was recently performing.
%
%
To enable edit recommendations based on temporal context, we present \technique, a novel
technique for learning edit sequence patterns from traces of developers' edits performed in an IDE.
%
%
Our experiments show that \technique has 78\% precision and
that \technique not only completed edits when developers missed the opportunity
to use the IDE tool support but also predicted new edits that have no tool
support in the IDE.

\end{abstract}

\maketitle

\section{Introduction}


Integrated Development Environments (IDEs) offer developers an overwhelming deluge of tools to support source code editing tasks, including writing new code, performing refactorings, and applying code fixes. Popular IDEs such as Microsoft Visual Studio~\cite{visualstudio} and JetBrains ReSharper~\cite{resharper}, for example, provide over 100 C\# refactorings, code fixes, and snippet tools.
Traditionally, these tools use the location where the developer is editing code and the surrounding code as \emph{spatial context} to generate candidate edits to recommend.

\begin{figure}
     \centering
     \begin{subfigure}[b]{0.404\textwidth}
         \centering
         \includegraphics[width=\textwidth]{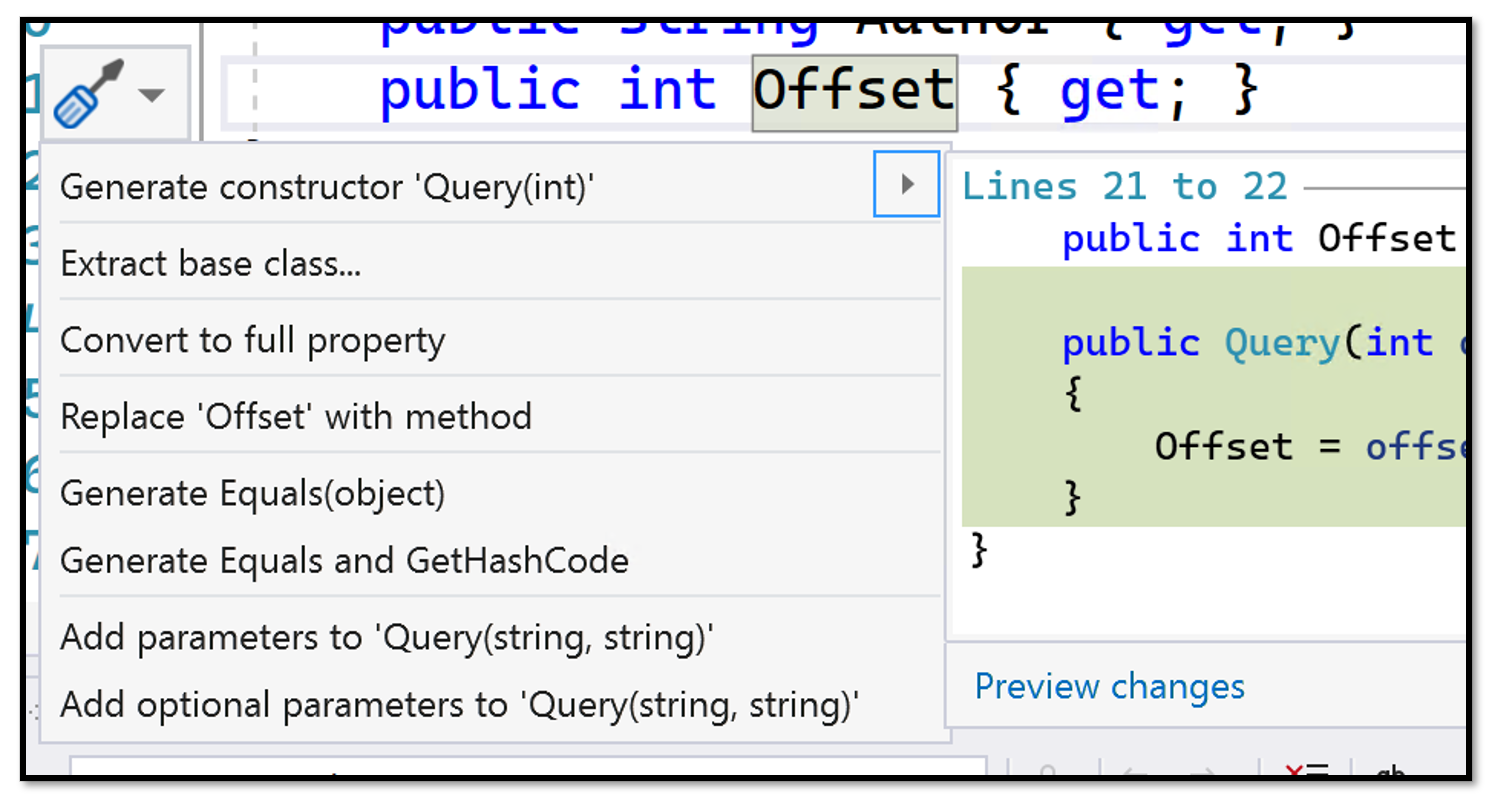}
         \caption{Edit suggestions based on spatial context}
         \label{fig:prop-menu}
     \end{subfigure}
     \begin{subfigure}[b]{0.54\textwidth}
         \centering
         \includegraphics[width=\textwidth]{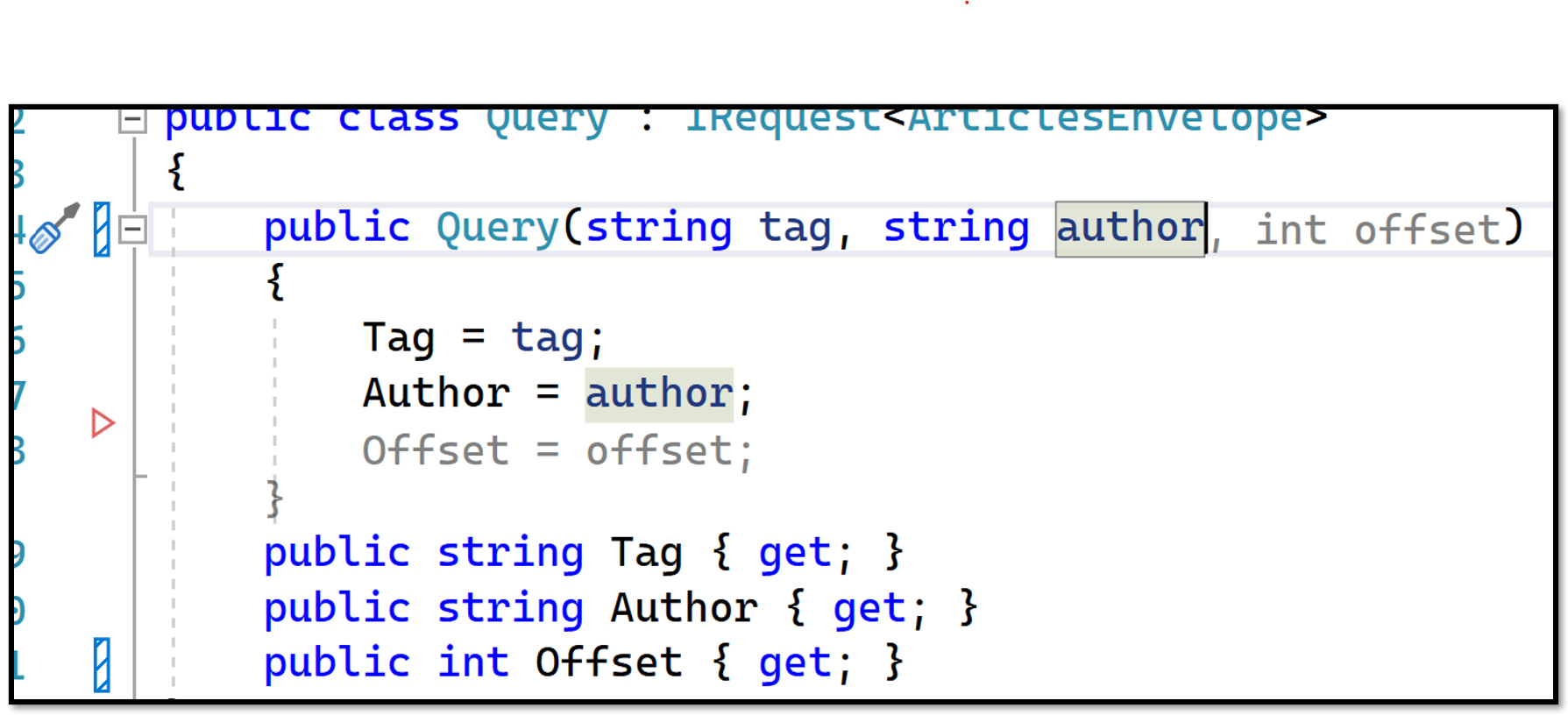}
         \caption{Edit suggestion using spatial and temporal contexts}
         \label{fig:add-prop-vs}
     \end{subfigure}
     \hfill
        \caption{Edits suggested by Visual Studio when the developer adds a property to a class}
        \label{fig:intro-example}
        \vspace{-4ex}
\end{figure}

However, the spatial context alone is often not sufficient for IDEs to confidently predict the developer’s next edit. At a specific location, there may be multiple candidate tools available for different editing tasks. For instance, Figure~\ref{fig:prop-menu} shows all tools available when the developer clicks on the screwdriver next to a property declaration. There are 8 edits that the IDE can automate at that location.  Unsurprisingly, developers have difficulty discovering these tools and applying them at the appropriate time and place~\cite{HowWeRefactor,Ge:ICSE:RefactoringSteps}.


To improve code edit recommendations,
in addition to the spatial context, we can also use the temporal context, that is,
the code edits that the developer was performing at a particular point in time.
For instance, suppose the developer has just added the \texttt{Offset} property in Figure~\ref{fig:prop-menu}. Next, the developer is more likely to add the corresponding parameter to the constructor and use it to initialize the property (7th option in the menu) than replace the nearly introduced property with a method (4th option). If the developer moves the cursor to the constructor, then it is very likely that they are about to insert the parameter. 
Recently, Visual Studio announced that they used this idea of temporal context to implement an analyzer to detect this \emph{edit sequence} and offer the suggestion as ``gray text'' (Figure~\ref{fig:add-prop-vs}) to add the parameter to the constructor as soon as the developer moves the cursor to the constructor after adding a new property.   
By using spatial and temporal contexts to generate suggestions at the right time and location, the IDE can afford to \emph{preemptively} show these edit suggestions, avoiding discoverability and late-awareness problems. 

However, implementing tools that use temporal context is non-trivial. Tool builders have to reason not only about the location where an edit should be suggested and how to automate the edit but also how previous edits relate to the edit under consideration. Consider the example above, developers can perform the "Insert Property", "Insert Parameter", "Insert Assignment" sequence of edits in any order but Visual Studio only handles the order shown in Figure~\ref{fig:intro-example}. 
Given the complexity of manually implementing these \emph{edit sequence patterns}, only a few of them are available today in IDEs. 



Instead of manually implementing patterns to recommend code edits, researchers have proposed several approaches to learn \emph{edit patterns} from edits in source code repositories~\cite{getafix, revisar, refazer, Kim:SystematicCodechange,yin2019learning}. These patterns represent the location where an edit should be applied and how to perform the desired edit. 
However, very few approaches use previous edits as temporal context. Blue-Pencil~\cite{Miltner:BluePencil} use previous edits to suggest similar repetitive edits. C$^3$PO~\cite{c3po} learns a model to complete an edit given a previous edit but uses only a single edit as context, and can only predict edits that do not generate new content. Additionally, C$^3$PO is trained on data from source code repositories, which may not reflect the temporal relationship of edits in an IDE.

In this paper, we propose \technique, a technique for learning \emph{Edit Sequence Patterns} from traces logged during editing sessions in the IDE.
As input, \technique takes a set of source file versions. Each version represents the state of the file while a developer is editing it. Given this input, \technique's problem is to find recurrent edit sequences and generalize them into Edit Sequence Patterns (ESPs). 
In a nutshell, \technique performs three major steps:
\begin{inparaenum}
  \item generating edit sequence sketches and their corresponding specifications; 
  \item synthesizing edit sequence patterns and
  \item selecting and ranking the edit sequence patterns.
\end{inparaenum}
Given a new development trace (i.e., edit history),
\technique can then use the learned edit sequence patterns
to predict the next edit.


To evaluate \technique, we collected $\mathsf{335,687}$ source file versions, which were logged from 12 professional software developers from a large company across several months.
%
In our experiments, \technique achieved $\mathsf{78.38\%}$ precision in the test set, showing a degree of domain-invariance, when compared to its performance on the validation set collected 6 months earlier. 
Additionally, we performed a qualitative analysis on the ESPs learned with \technique. Our findings show that ESPs can be used not only to complete edits when developers typically miss the opportunity to use the IDE tool support but also to predict new edits that have no tool support at all in the IDE.

%
%
%

In short, the paper makes the following contributions: 
\begin{enumerate}
    \item We formalize the problem of learning  Edit Sequence Patterns (ESPs) (Section~\ref{sec:prelims});
    \item  We propose \technique, a technique for learning edit sequences patterns from traces collected during editing sessions in the IDE (Sections~\ref{sec:generate_sketch}-\ref{sec:rank_esp});
    \item We show that the ESPs learned by \technique can be used to predict edits with 78.38\% precision (Section~\ref{sec:results});  
    \item Our qualitative analysis shows that ESPs can be used not only to complete edits when developers missed the opportunity to use the IDE tool support but also predict new edits that have no tool support at all in the IDE (Section~\ref{sec:results}).
\end{enumerate}

\section{Overview}
\label{sec:motivating-example}

We begin with an overview of \technique's technique to learn ESPs and how we can use these patterns to predict code edits. To illustrate the process, we show how \technique learns an ESP that predicts the code edit recommended by Visual Studio in Figure~\ref{fig:add-prop-vs}. As we mentioned, Visual Studio developers had to manually implement this feature, which is time-consuming and hard to scale. In Section~\ref{sec:results} we present a list of other patterns that were automatically learned by \technique.

\begin{figure}
  \scriptsize
%
\hfill
\begin{subfigure}{0.30\textwidth}
\begin{lstlisting}[
  language=diff,
  basicstyle=\ttfamily\scriptsize
]
  class Node {
  
    Node() {
    }
  }
\end{lstlisting}
\vspace{-6ex}
  \caption{$\version_0$}
\end{subfigure}
\hfill
\begin{subfigure}{0.30\textwidth}
\begin{lstlisting}[
  language=diff,
  basicstyle=\ttfamily\scriptsize,
  breaklines=false
]
  class Node {
+   public str Id { }
    Node() {
    }
  }
\end{lstlisting}
\vspace{-6ex}
  \caption{$\version_1$}
\end{subfigure}
\hfill
%
%
%
\begin{subfigure}{0.30\textwidth}
\begin{lstlisting}[
  language=diff,
  basicstyle=\ttfamily\scriptsize,
  breaklines=false
]
  class Node {
-   public str Id { }
+   public str Id {get;}
    Node() {
    }
  }
\end{lstlisting}
\vspace{-6ex}
  \caption{$\version_2$}
\end{subfigure}
\hfill

\hfill
\begin{subfigure}{0.30\textwidth}
  \begin{lstlisting}[
  language=diff,
  basicstyle=\ttfamily\scriptsize,
  breaklines=false
]
  class Node {
-   public str Id {get;}
+   public str Id {get;set;}
    Node() {
    }
  }
  \end{lstlisting}
\vspace{-6ex}
  \caption{$\version_3$}
\end{subfigure}
\hfill
%
%
%
\begin{subfigure}{0.30\textwidth}
  \begin{lstlisting}[
  mathescape,
  escapeinside={(*}{*)},
  language=diff,
  basicstyle=\ttfamily\scriptsize,
  breaklines=false
]
  class Node {
    public str Id {get;set;}
-   Node() {
+   Node(str id) {
 
    }
  }
  \end{lstlisting}
\vspace{-6ex}
  \caption{$\version_4$}
\end{subfigure}
\hfill
\begin{subfigure}{0.30\textwidth}
  \begin{lstlisting}[
  language=diff,
  basicstyle=\ttfamily\scriptsize,
  breaklines=false
]
  class Node {
    public str Id {get;set;}
    Node(str id) {
+     Id = id;
    }
  }
  \end{lstlisting}
\hfill
\vspace{-6ex}
  \caption{$\version_5$}
\end{subfigure}
\vspace{-2ex}
  \caption{Development Session: Syntactically correct versions
  while adding and initializing a property.}
  \label{fig:dev-session}
\vspace{-5ex}
\end{figure}

\begin{figure}
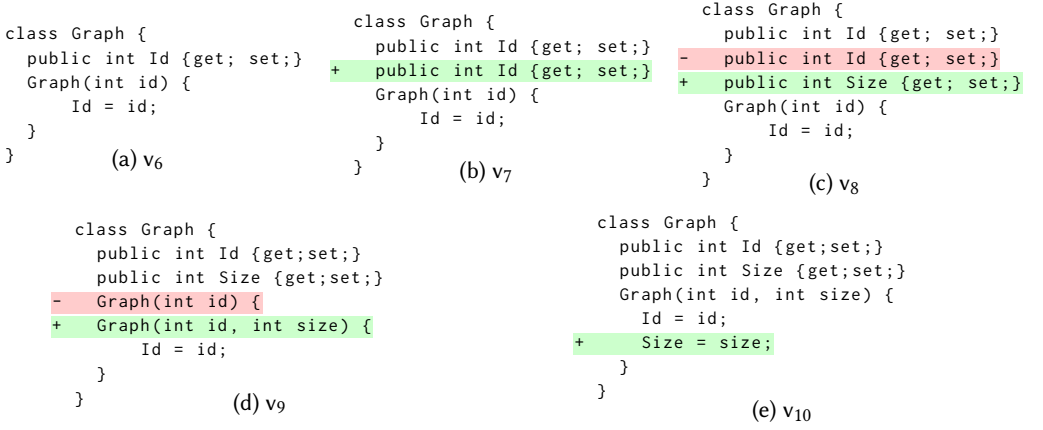

  \scriptsize
%
\hfill
\begin{subfigure}{0.30\textwidth}
\begin{lstlisting}[
  language=diff,
  basicstyle=\ttfamily\scriptsize
]
  class Graph {
    public int Id {get; set;}
    Graph(int id) {
        Id = id;
    }
  }
\end{lstlisting}
\vspace{-6ex}
  \caption{$\version_6$}
\end{subfigure}
\hfill
\begin{subfigure}{0.30\textwidth}
\begin{lstlisting}[
  language=diff,
  basicstyle=\ttfamily\scriptsize,
  breaklines=false
]
  class Graph {
    public int Id {get; set;}
+   public int Id {get; set;}
    Graph(int id) {
        Id = id;
    }
  }
\end{lstlisting}
\vspace{-6ex}
  \caption{$\version_7$}
\end{subfigure}
\hfill
%
%
%
\begin{subfigure}{0.30\textwidth}
\begin{lstlisting}[
  language=diff,
  basicstyle=\ttfamily\scriptsize,
  breaklines=false,
]
  class Graph {
    public int Id {get; set;}
-   public int Id {get; set;}
+   public int Size {get; set;}
    Graph(int id) {
        Id = id;
    }
  }
\end{lstlisting}
\vspace{-6ex}
  \caption{$\version_8$}
\end{subfigure}
\hfill

%
%
%
 \hfill
\begin{subfigure}{0.40\textwidth}
  \begin{lstlisting}[
  mathescape,
  language=diff,
  basicstyle=\ttfamily\scriptsize,
  breaklines=false
]
  class Graph {
    public int Id {get;set;}
    public int Size {get;set;}
-   Graph(int id) { 
+   Graph(int id, int size) {
        Id = id;
    }
  }
  \end{lstlisting}
\vspace{-6ex}
  \caption{$\version_{9}$}
\end{subfigure}
\hfill
\begin{subfigure}{0.40\textwidth}
  \begin{lstlisting}[
  language=diff,
  basicstyle=\ttfamily\scriptsize,
  breaklines=false
]
  class Graph {
    public int Id {get;set;}
    public int Size {get;set;}
    Graph(int id, int size) {
      Id = id;
+     Size = size;
    }
  }
  \end{lstlisting}
\hfill
\vspace{-6ex}
  \caption{$\version_{10}$}
\end{subfigure}
\vspace{-2ex}
  \caption{Development Session: Syntactically correct versions
  while copying, updating, and initializing a property. 
  } 
  \label{fig:dev-session2}
 \vspace{-2ex}
\end{figure}

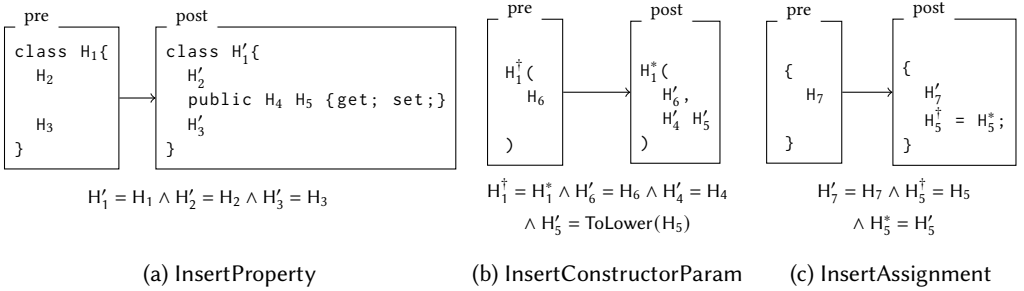
\begin{figure}[t!]
\begin{subfigure}[t]{0.45\textwidth}
\centering
\begin{tikzpicture}[main/.style = {draw, circle}]
\tikzstyle{box}=[
      draw,
      rectangle,
      minimum width=2.00cm,
      minimum height=0.75cm,
    ]
\node[draw] (before) {
    \begin{lstlisting}[
          mathescape,
          basicstyle=\ttfamily\scriptsize,
        ]
class H$_1${
  H$_2$
  
  H$_3$
}
    \end{lstlisting}};
\node[xshift=1ex,yshift=-0.7ex, overlay, fill=white, draw=white, above 
right] at (before.north west) {
\scriptsize pre
};    
\node[draw,right of=before, xshift=2.25cm] (after) {
    \begin{lstlisting}[
        mathescape,
        language=diff,
        basicstyle=\ttfamily\scriptsize,
        ]
class H$_1'${
  H$_2'$
  public H$_4$ H$_5$ {get; set;}
  H$_3'$
}
    \end{lstlisting}
};
\node[yshift=-0.5cm, xshift=-1.3cm, below of=after][align=center](tmprel1) {\scriptsize
  {\scriptsize $\mathsf{\hole_1' = \hole_1 \land \hole_2' = \hole_2 \land
  \hole_3' = \hole_3}$} \\
  {\scriptsize }
};
    \node[xshift=1ex,yshift=-0.7ex, overlay, fill=white, draw=white, above 
right] at (after.north west) {
\scriptsize post
};   
\draw[->](before) -- (after); 
\end{tikzpicture}
\caption{$
      \mathsf{InsertProperty}$}
\label{fig:pattern-example-a}
\end{subfigure}
\begin{subfigure}[t]{0.26\textwidth}
\centering
\begin{tikzpicture}[main/.style = {draw, circle}]
\tikzstyle{box}=[
      draw,
      rectangle,
      minimum width=1.00cm,
      minimum height=0.75cm,
    ]
\node[box] (before) {
    \begin{lstlisting}[
            mathescape,
          basicstyle=\ttfamily\scriptsize,
        ]
        
H$_1^\dagger$(
  H$_6$
    
)
    \end{lstlisting}};
 \node[xshift=1ex,yshift=-0.7ex, overlay, fill=white, draw=white, above 
right] at (before.north west) {
\scriptsize pre
};     
\node[box,right of=before, xshift=1cm] (after) {
    \begin{lstlisting}[
        mathescape,
        language=diff,
        basicstyle=\ttfamily\scriptsize,
        breaklines=false,
        ]
        
H$_1^*$(
  H$_6'$,
  H$_4'$ H$_5'$
)
    \end{lstlisting}};
\node[yshift=-0.5cm, xshift=-0.9cm, below of=after][align=center](tmprel2) {\scriptsize
{\scriptsize$\mathsf{\hole_1^\dagger = \hole_1^* \land \hole_6' = \hole_6  \land \hole_4' = \hole_4}$} \\ 
{\scriptsize $\mathsf{ \land~\hole_5' = ToLower(\hole_5)}$}
};
\node[xshift=1ex,yshift=-0.7ex, overlay, fill=white, draw=white, above 
right] at (after.north west) {
\scriptsize post
};  
\draw[->](before) -- (after); 
\end{tikzpicture}
\caption{$
      \mathsf{InsertConstructorParam}$}
\label{fig:pattern-example-b}
\end{subfigure}
\begin{subfigure}[t]{0.27\textwidth}
\centering
\begin{tikzpicture}[main/.style = {draw, circle}]
\tikzstyle{box}=[
      draw,
      rectangle,
      minimum width=1.00cm,
      minimum height=0.75cm,
    ]
\node[box] (before) {
    \begin{lstlisting}[
            mathescape,
          basicstyle=\ttfamily\scriptsize,
        ]
        
{
  H$_{7}$
  
}
    \end{lstlisting}};
 \node[xshift=1ex,yshift=-0.7ex, overlay, fill=white, draw=white, above 
right] at (before.north west) {
\scriptsize pre
};    
\node[box,right of=before, xshift=1cm] (after) {
    \begin{lstlisting}[
        mathescape,
        language=diff,
        basicstyle=\ttfamily\scriptsize,
        breaklines=false,
        ]
        
{
  H$_{7}'$
  H$_5^\dagger$ = H$_{5}^*$;
}
    \end{lstlisting}};
    \node[xshift=1ex,yshift=-0.7ex, overlay, fill=white, draw=white, above 
right] at (after.north west) {
\scriptsize post
};

\node[yshift=-0.5cm, xshift=-0.8cm, below of=after][align=center](tmprel3) {\scriptsize
{\scriptsize $\mathsf{\hole_7' = \hole_7 \land H_5^\dagger = \hole_5}$} \\ 
{\scriptsize $\mathsf{ \land~H_{5}^* = \hole_5'}$}
};
\draw[->](before) -- (after); 
\end{tikzpicture}
\caption{$
      \mathsf{InsertAssignment}$}
\label{fig:pattern-example-c}
\end{subfigure}
\vspace{-2ex}
\caption{Example of an Edit Sequence Pattern learned by \technique for the workflow $
      \mathsf{InsertProperty}
\cdot \mathsf{InsertConstructorParameter}
\cdot \mathsf{InsertAssignment}$ The variable component of the pattern (holes) are represented by $\hole$. Below each \emph{pre} and \emph{post} representaion of the template, we present the \emph{Hole Predicates} specifying the relationship between holes across the edit pattern sequence.}
\label{fig:pattern}
\vspace{-3ex}
\end{figure}
Consider the source file traces shown in Figures~\ref{fig:dev-session} and~\ref{fig:dev-session2} depicting the sequences of versions produced when developers were performing similar edits in an IDE.
At a high level, the developers are performing the same ESP:
\begin{inparaenum}[(a)]
  \item adding a new property to the class,
  \item adding a new parameter to the constructor with the same name as of the property (but lowercase) and same type,
  \item adding a statement assigning the parameter to the property.
\end{inparaenum}
However, the developers take different paths in the two cases---in Figure~\ref{fig:dev-session}, the developer directly types in the new code while in Figure~\ref{fig:dev-session2}, the developer copies an existing property and changes the name.
Figure~\ref{fig:pattern} illustrates how \technique represents this pattern. Each individual transition represents the \emph{pre} and \emph{post} template of an edit template. We see that the \emph{insert property} pattern in Figure~\ref{fig:pattern-example-a} has templates with holes in it. Holes $\hole_2$ and $\hole_3$, respectively, represent the surrounding class members and methods preceding and following the location of the edit. The \textit{type} and \textit{name} of the added property in the \emph{post} template correspond to holes $\hole_4$ and $\hole_5$, respectively.
Based on the newly added property, holes $\hole_4$ and $\hole_5$ can be
replaced with the appropriate type and name to match the edit.

We use hole predicates to define relationships between holes in the pre- and
post-templates.
The predicates $\hole_i' = \hole_i$  for $i \in \{ 1, 2, 3 \}$ represent that
the class name and the class body does not change apart from the newly added
property.
Similarly, in
Figure~\ref{fig:pattern-example-b}, the predicate $\hole_6' = \hole_6$ represents that the constructor parameters do not change except the newly added parameter.
The predicate $\hole_5' = \mathsf{ToLower}(\hole_5)$ says that the name of the
parameter is the lower case version of the property name (e.g., if the property
name is \code{Id}, the parameter name will be \code{id}).
Note that this predicate relates the holes in two different edit templates,
i.e., $\hole_5$ is in the $\mathsf{InsertProperty}$ template while $\hole_5'$ is in
$\mathsf{InsertConstructorParam}$.
Hence, while learning an ESP, we need to consider the sequence
of edits as a whole, instead of separately learning single edit patterns and
putting them together.

\subsection{Using Edit Sequence Patterns to Predict Edits}
\label{sec:using-patterns}
\ignore{
We can use the the above pattern to predict the next edits that the developer will perform. For instance, consider the scenario shown in Figure~\ref{fig:dev-session}. Suppose the developer has just performed the changes $\version_0 \to \version_3$ and moved the cursor location to the parameter list to make the change $\version_3 \to \version_4$. At this point, we can match the edit $\version_0 \to \version_3$ to the first edit template in our ESP, and we can match the parameter list where the cursor is placed to the left-hand side template of the second edit template. We then use the right-hand side template to produce the edit. Note that we know how to fill in the holes on the right-hand side by using the name and type of the added property. Similarly, we can suggest the edit that adds the assignment after that. 
The same ESP also applies to $\version_8 \to \version_9$ and $\version_9 \to \version_{10}$ after the change $\version_6 \to \version_8$ in Figure~\ref{fig:dev-session2}.
\mohammad{to complete the illustration, would be good to quickly mention that the different sequence of edits in Figure 3 are also captured by the same general pattern (v6->v8, v8->v9, etc).} \yh{added.}
}
We can use the above pattern to predict the next edits that the developer will
perform.
For instance, consider the scenario shown in Figure~\ref{fig:dev-session}.
Suppose the developer has just performed the changes $\version_0 \to \version_3$.
We can match this edit to $\mathsf{InsertProperty}$ to get the values of the holes
$\hole_4$ and $\hole_5$, i.e., \code{str} and \code{Id}, respectively.
Now, using the predicates $\hole_4' = \hole_4$ and $\hole_5' =
\mathsf{ToLower}(\hole_5)$, we can instantiate $\mathsf{InsertConstructorParam}$
to obtain the next edit.
In an IDE, we can use this instantiation to suggest adding \code{str id} as soon
as the developer moves the cursor to the constructor's parameter list using an
interface similar to the one shown in Figure~\ref{fig:add-prop-vs}.
Note that in Figure~\ref{fig:add-prop-vs}, we can
predict two subsequent changes (adding the constructor parameter and adding an assignment) at once using edit patterns $\mathsf{InsertConstructorParam}$ and $\mathsf{InsertAssignment}$ in sequence.
%
%
%
%
%

\ignore{
Note, however, that there may be alternative edits that a developer may want to make at a given point in time. For instance, Figure~\ref{fig:dev-session3} shows the trace of a developer performing a similar sequence of edits. Here, after adding a property to the class, the developer adds a parameter of a different type and name to the constructor \mohammad{only the name is different, type is int for both cases} \yasharth{Made some changes to the example as having just the name different was not very convincing of a different workflow}. The code edit produced with the previous ESP would not be appropriate for this developer at this point in time. That is why IDEs present these edits as recommendations or suggestions instead of automatically editing the developer's code. When a recommendation is not valid, the developer can quickly skip it by pressing ``esc'' or typing something different. Additionally, after adding the parameter \mohammad{is this an incomplete sentence?}.  
}

Note that the predictions made using the ESP is just that, a
prediction.
As shown in Figure~\ref{fig:dev-session3}, the developer may actually want to
make a different sequence of changes, i.e., the name and type of the property
and the initialization expression are different that the ones predicted by the
ESP.
In our IDE plugin implementation, the developer can press the Escape key to
ignore the recommendation from the edit sequence template and make their own change.
%
\begin{figure}
  \scriptsize
%
\hfill
\begin{subfigure}{0.30\textwidth}
\begin{lstlisting}[
  language=diff,
  basicstyle=\ttfamily\scriptsize
]
  class Metric {
  
    Metric() {
    }
  }
\end{lstlisting}
\vspace{-6ex}
  \caption{$\version_{11}$}
\end{subfigure}
\hfill
\begin{subfigure}{0.30\textwidth}
\begin{lstlisting}[
  language=diff,
  basicstyle=\ttfamily\scriptsize,
  breaklines=false
]
 class Metric {
+  public float Cost { }
   Metric() {
   }
 }
\end{lstlisting}
\vspace{-6ex}
  \caption{$\version_{12}$}
\end{subfigure}
\hfill
%
%
%
\begin{subfigure}{0.30\textwidth}
\begin{lstlisting}[
  language=diff,
  basicstyle=\ttfamily\scriptsize,
  breaklines=false
]
 class Metric {
-  public float Cost { }
+  public float Cost {get;}
   Metric() {
   }
 }
\end{lstlisting}
\vspace{-6ex}
  \caption{$\version_{13}$}
\end{subfigure}
\hfill

\hfill
\begin{subfigure}{0.34\textwidth}
  \begin{lstlisting}[
  language=diff,
  basicstyle=\ttfamily\scriptsize,
  breaklines=false
]
 class Metric {
-  public float Cost {get;}
+  public float Cost {get;set;}
   Metric() {
   }
 }
  \end{lstlisting}
\vspace{-6ex}
  \caption{$\version_{14}$}
\end{subfigure}
\hfill
%
%
%
\begin{subfigure}{0.33\textwidth}
  \begin{lstlisting}[
  mathescape,
  language=diff,
  basicstyle=\ttfamily\scriptsize,
  breaklines=false
]
 class Metric {
   public float Cost {get;set;}
-  Metric() {
+  Metric(int val) {
 
   }
 }
  \end{lstlisting}
\vspace{-6ex}
  \caption{$\version_{15}$}
\end{subfigure}
\hfill
\begin{subfigure}{0.32\textwidth}
  \begin{lstlisting}[
  language=diff,
  basicstyle=\ttfamily\scriptsize,
  breaklines=false
]
 class Metric {
   public float Cost {get;set;}
   Metric(int val) {
+    Cost = Math.Abs(val);
   }
 }
  \end{lstlisting}
\hfill
\vspace{-6ex}
  \caption{$\version_{16}$}
\end{subfigure}
\vspace{-2ex}
  \caption{Another sequence of versions that is different from the edit sequence pattern learned in Fig~\ref{fig:pattern}.}
  \label{fig:dev-session3}
\vspace{-3ex}
\end{figure}

\subsection{Learning Edit Sequence Patterns}

\ignore{Given the traces presented in Figures~\ref{fig:dev-session}, ~\ref{fig:dev-session2}, and ~\ref{fig:dev-session3} as input, \technique first creates the edit graph in Figure~\ref{fig:edit-graph} to represent edits at different levels of granularity and their temporal relationship \mohammad{what is meant by "a different level of granularity" - different than what?}\priyanshu{Hope it looks better now} \mohammad{im not clear - is the initial edit graph created a complete graph of all possible vi-vj where i < j ? If not then how do we choose which pairs of edits to keep and which not?\ The edges represent temporal sequence right? }\priyanshu{Check with Gustavo} \yh{I am not sure the temporal relationship aspect neither}. The node labels show the pair versions that constitute an edit. For instance, the nodes $\version_0 \to \version_1$, $\version_1 \to \version_2$, and $\version_2 \to \version_3$,  represent the finest-granularity edits that add the \texttt{Id} property with an empty body, the get accessor, and the set accessor, respectively. The node $\version_0 \to \version_3$ represents a coarser-granularity edit that inserts this property and its get and set accessors. 
} 
\ignore{However, there is only one edge between $\version_7\version_8$ and $\version_8\version_9$ in the second edit graph.}

Given the traces presented in Figures~\ref{fig:dev-session},
~\ref{fig:dev-session2}, and ~\ref{fig:dev-session3} as input, the aim of
\technique is to learn the ESP in Figure~\ref{fig:pattern}.
%
%

\noindent\textbf{Building the Edit Graph.}
\technique first creates the edit graph in Figure~\ref{fig:edit-graph} to
where the nodes represent edits at different \emph{levels of granularity} and the edges
represent their temporal relationship.
For example, the figure contains both the node $\version_0 \to \version_3$, as
well as the nodes $\version_0 \to \version_1$, $\version_1 \to \version_2$, and
$\version_2 \to \version_3$; these represent the same change of adding the
property \code{public str Id \{ get; set; \}}, but at different levels of
granularity.
The edit $\version_0 \to \version_3$ represents adding the full property, while
$\version_0 \to \version_1$, $\version_1 \to \version_2$, and
$\version_2 \to \version_3$ represent adding the property with the empty
accessor list, adding the \code{get;}, and adding the \code{set;}.
The edges between $\version_0 \to \version_1$, $\version_1 \to \version_2$, and
$\version_2 \to \version_3$ represent that each edit immediately follows the
previous in the trace.
Note that the graph does not contain nodes for all changes (for example,
$\version_0 \to \version_5$).
We describe how we select the edits that should be there in the graph in
Section~\ref{sec:generate_sketch}--intuitively, we ignore large and unrelated edits.
\ignore{
\arsays{Compare the previous paragraph to this one and check which one reads better.}
 \mohammad{what
is meant by "a different level of granularity" - different than
what?}\priyanshu{Hope it looks better now} \mohammad{im not clear - is the
initial edit graph created a complete graph of all possible vi-vj where i < j ?
If not then how do we choose which pairs of edits to keep and which not?\ The
edges represent temporal sequence right? }\priyanshu{Check with Gustavo}. The
node labels show the pair versions that constitute an edit. For instance, the
nodes $\version_0 \to \version_1$, $\version_1 \to \version_2$, and $\version_2
\to \version_3$,  represent the edits that add the \texttt{Id} property with an
empty body, the get accessor, and the set accessor, respectively. The node
$\version_0 \to \version_3$ represents the larger edit that inserts this
property and its get and set accessors. 
}

\noindent\textbf{Creating Sketches for Edit Pattern Sequences.}
Next, \technique produces a quotient graph by grouping together similar
edits in the edit graph.
Two edits are grouped together, i.e., in the same partition, if they have the
same edit type (Insert, Delete, or Update) and the same type of AST node that is
being \ignore{deleted, inserted, or updated} modified 
(e.g., $\mathsf{PropertyDeclaration}$ and
$\mathsf{Parameter}$).
In Figure~\ref{fig:edit-graph}, the nodes are colored by partition.
For example, the green nodes all represent the insertion of a $\mathsf{PropertyDeclaration}$.

Figure~\ref{fig:quotient-graph} shows the quotient graph produced by \technique.
The \emph{quotient graph} summarizes the edit graph at the level of partitions:
the vertices of the quotient graph are the partitions.
An edge between two partitions exists in the quotient graph \ignore{if and only if} \emph{iff} there
are at least $2$ pairs of edits in the partitions that sequentially follow each other.
For example, there is an edge between $\mathsf{InsertProperty}$ and
$\mathsf{InsertParameter}$ as there are $3$ pairs of edits where a parameter is
added immediately after a property is added (see edge label in Figure~\ref{fig:quotient-graph}).
However, there are no edges to and from \ignore{the partition} $\mathsf{UpdateName}$
since no two occurrences of update name are followed by  edits of the same partition.
%
%

\ignore{
Each quotient graph edge contains a set of pairs of edits that have corresponding edges in the edit graphs. 
There is an edge between two partitions if there are at least two edges between
the edits in the partitions.

Figure~\ref{fig:quotient-graph} shows the quotient graph produced by \technique.
The quotient graph vertices are partitions of edits with the same edit type and
edited AST node type. 
Each quotient graph edge contains a set of pairs of edits that have corresponding edges in the edit graphs. 
Note that there are no in- and out-edges of node ``UpdateName'' because we
construct an edge in quotient graph if there are at least $\edseqnumber = 2$
different pairs of edits.
}

\ignore{
\begin{figure}
     \centering
     \begin{subfigure}[b]{\textwidth}
         \centering
         \includegraphics[width=\textwidth]{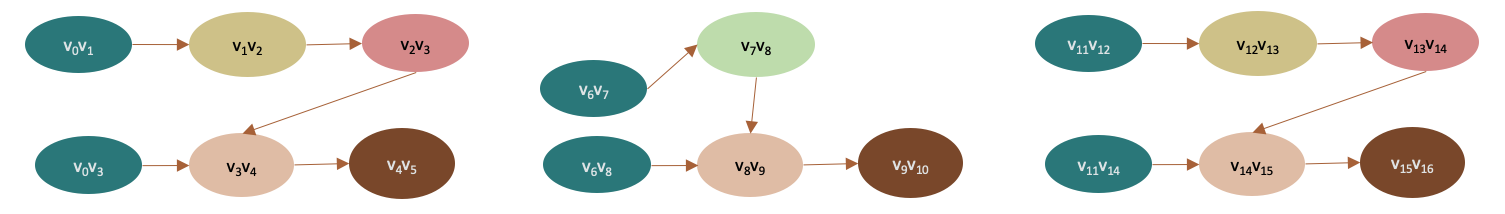}
         \caption{Edit graph. Notice that $\version_0\version_2$ and $\version_{11}\version_{13}$ are also ``InsertProp'' nodes, but we omit them for ease of presentation.}
         \label{fig:edit-graph}
     \end{subfigure}
     \hfill
     \begin{subfigure}[b]{\textwidth}
         \centering
         \includegraphics[width=0.6\textwidth]{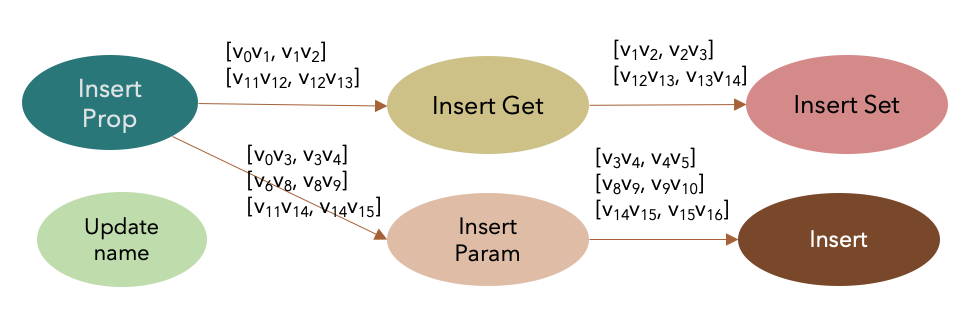}
         \arsays{Insert should be InsertAssignment}
         \caption{Quotient graph.}
         \label{fig:quotient-graph}
     \end{subfigure}
     \begin{subfigure}[c]{\textwidth}
         \centering
         \includegraphics[width=\textwidth]{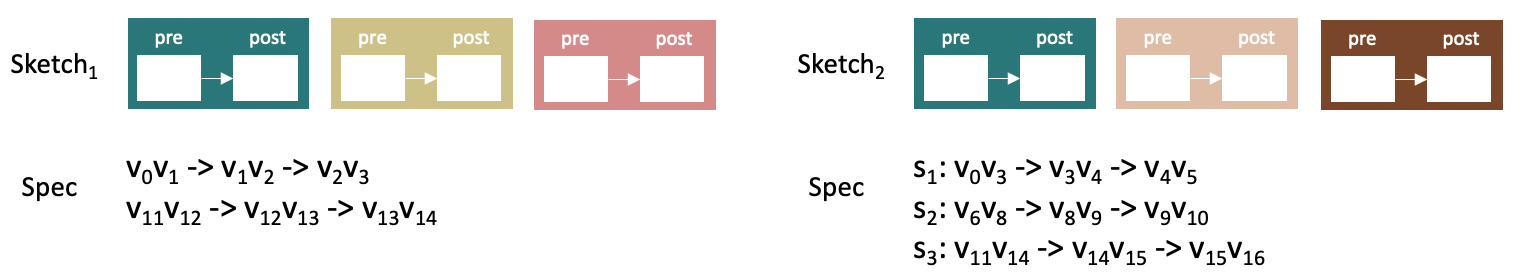}
         \caption{Sketches of the ESP. On the left: ``InsertProp''$\to$``InsertGet''$\to$``InsertSet'', and on the right: ``InsertProp''$\to$``InsertParam''$\to$``Insert''.}
         \label{fig:edit-pattern-sketch}
     \end{subfigure}
     \caption{\technique: From Edit Graphs to Edit Sequence Patterns}
\end{figure}
}
\begin{figure}
\begin{subfigure}{\textwidth}
\centering
\scalebox{0.8}{
\begin{tikzpicture}
\tikzstyle{box}=[
  draw,
  rectangle
]
\tikzstyle{insertprop}=[box, fill=teal!60]
\tikzstyle{insertget}=[box, fill=olive!30]
\tikzstyle{insertset}=[box, fill=magenta!30]
\tikzstyle{insertparam}=[box, fill=orange!20]
\tikzstyle{insertassign}=[box, fill=brown!80]
\tikzstyle{updatename}=[box, fill=green!15]

  \node[insertprop]                              (v0v1) {$\version_0 \to \version_1$};
  \node[insertget,    right of=v0v1, xshift=1.0cm] (v1v2) {$\version_1 \to \version_2$};
  \node[insertset,    right of=v1v2, xshift=1.0cm] (v2v3) {$\version_2 \to \version_3$};
  \node[insertprop,   below of=v0v1, yshift=0.0cm] (v0v3) {$\version_0 \to \version_3$};
  \node[insertparam,  below of=v1v2, yshift=0.0cm] (v3v4) {$\version_3 \to \version_4$};
  \node[insertassign, below of=v2v3, yshift=0.0cm] (v4v5) {$\version_4 \to \version_5$};

  \draw[->] (v0v1) -- (v1v2);
  \draw[->] (v1v2) -- (v2v3);
  \draw[->] (v0v3) -- (v3v4);
  \draw[->] (v3v4) -- (v4v5);
  \draw[->] (v2v3) -- ++(0cm, -0.5cm) -| (v3v4);

  \node[insertprop,   right of=v2v3, xshift=1.0cm] (v6v7)  {$\version_6 \to \version_7$};
  \node[insertprop,   below of=v6v7, yshift=0.0cm] (v6v8)  {$\version_6 \to \version_8$};
  \node[updatename,   right of=v6v7, xshift=1.0cm] (v7v8)  {$\version_7 \to \version_8$};
  \node[insertparam,  right of=v6v8, xshift=1.0cm] (v8v9)  {$\version_8 \to \version_9$};
  \node[insertassign, right of=v8v9, xshift=1.0cm] (v9v10) {$\version_9 \to \version_{10}$};

  \draw[->] (v6v7) -- (v7v8);
  \draw[->] (v7v8) -- (v8v9);
  \draw[->] (v6v8) -- (v8v9);
  \draw[->] (v8v9) -- (v9v10);

  \node[insertprop,   right of=v7v8,   xshift=3.0cm]  (v11v12) {$\version_{11} \to \version_{12}$};
  \node[insertget,    right of=v11v12, xshift=1.2cm]  (v12v13) {$\version_{12} \to \version_{13}$};
  \node[insertset,    right of=v12v13, xshift=1.2cm]  (v13v14) {$\version_{13} \to \version_{14}$};
  \node[insertprop,   below of=v11v12, yshift=0.0cm]  (v11v14) {$\version_{11} \to \version_{14}$};
  \node[insertparam,  below of=v12v13, yshift=0.0cm]  (v14v15) {$\version_{14} \to \version_{15}$};
  \node[insertassign, below of=v13v14, yshift=0.0cm]  (v15v16) {$\version_{15} \to \version_{16}$};

  \draw[->] (v11v12) -- (v12v13);
  \draw[->] (v12v13) -- (v13v14);
  \draw[->] (v11v14) -- (v14v15);
  \draw[->] (v14v15) -- (v15v16);
  \draw[->] (v13v14) -- ++(0cm, -0.5cm) -| (v14v15);
\end{tikzpicture}
}
\caption{Part of edit graph for traces from Figures~\ref{fig:dev-session},~\ref{fig:dev-session2}, and~\ref{fig:dev-session3}}
\label{fig:edit-graph}
\end{subfigure}

\vspace{2ex}
\begin{subfigure}{\textwidth}
\centering
\scalebox{0.7}{
\begin{tikzpicture}
\tikzstyle{partition}=[
  draw,
  rectangle,
  minimum width=4.00cm
]
\tikzstyle{insertprop}=[partition, fill=teal!60]
\tikzstyle{insertget}=[partition, fill=olive!30]
\tikzstyle{insertset}=[partition, fill=magenta!30]
\tikzstyle{insertparam}=[partition, fill=orange!20]
\tikzstyle{insertassign}=[partition, fill=brown!80]
\tikzstyle{updatename}=[partition, fill=green!15]

\node[insertprop] (addprop) {
  \tabular{c}
    Insert Property\\
    \hline
    $\version_0 \to \version_1$, $\version_0 \to \version_3$ \\
    $\version_6 \to \version_7$, $\version_6 \to \version_8$ \\
    $\version_{11} \to \version_{12}$, $\version_{11} \to \version_{14}$ \\
  \endtabular
};
\node[insertget, right=of addprop.north east, anchor=north, xshift=4cm] (addget) {
  \tabular{c}
    Insert Get\\
    \hline
    $\version_1 \to \version_2$, $\version_{12} \to \version_{13}$\\
  \endtabular
};
\node[updatename, below of=addprop, yshift=-2cm] (upname) {
  \tabular{c}
    Update Name\\
    \hline
    $\version_7 \to \version_8$
  \endtabular
};
\node[insertparam, right=of upname.south east, anchor=south, xshift=4cm] (addparam) {
  \tabular{c}
    Insert Parameter\\
    \hline
    $\version_3 \to \version_4$, $\version_8 \to \version_9$\\
    $\version_{14} \to \version_{15}$
  \endtabular
};
\node[insertset, right=of addget.north east, anchor=north, xshift=4cm] (addset) {
  \tabular{c}
    Insert Set\\
    \hline
    $\version_2 \to \version_3$, $\version_{13} \to \version_{14}$\\
  \endtabular
};
\node[insertassign, right=of addparam.south east, anchor=south, xshift=4cm] (addassign) {
  \tabular{c}
    Insert Assignment\\
    \hline
    $\version_4 \to \version_5$, $\version_9 \to \version_{10}$\\
    $\version_{15} \to \version_{16}$\\
  \endtabular
};

\draw[->] (addprop.east|-addget.west) --  node  {{\small
  \tabular{c}
  $\version_0 \to \version_1 \to \version_2$\\
  $\version_{11} \to \version_{12} \to \version_{13}$\\
  \endtabular
}} (addget);

\draw[->] (addprop.south)
          -- ++(0.0cm, -0.75cm)
          -| node [below, near start, yshift=0.85cm] {{\small
            \tabular{c}
            $\version_0 \to \version_1 \to \version_2$\\
            $\version_6 \to \version_8 \to \version_9$\\
            $\version_{11} \to \version_{12} \to \version_{13}$\\
            \endtabular
          }}
          ($(addparam.west)-(0.5cm,0cm)$)
          -- (addparam.west);

\draw[->] (addget.east) -- node {{\small
  \tabular{c}
  $\version_1 \to \version_2 \to \version_3$\\
  $\version_{12} \to \version_{13} \to \version_{14}$\\
  \endtabular
}} (addset.west);

\draw[->] (addset.south)
          -- ++(0.0cm, -0.75cm)
          -| node [near start] {{
            \tabular{c}
            $\version_2 \to \version_3 \to \version_4$\\
            $\version_{13} \to \version_{14} \to \version_{15}$
            \endtabular
          }}
          (addparam.north);

\draw[->] (addparam.east) -- node[below, yshift=0.55cm] {{
  \tabular{c}
  $\version_3 \to \version_4 \to \version_5$\\
  $\version_8 \to \version_9 \to \version_{10}$\\
  $\version_{14} \to \version_{15} \to \version_{16}$\\
  \endtabular
}} (addassign.west);

\end{tikzpicture}
}
\caption{Quotient graph for edit graph in Figure~\ref{fig:edit-graph}}
\label{fig:quotient-graph}
\end{subfigure}

\vspace{2ex}
\begin{subfigure}{\textwidth}
\begin{tikzpicture}
\tikzstyle{template}=[
  draw,
  fill=white,
  rectangle,
  minimum width=0.50cm,
  minimum height=0.50cm
]
\tikzstyle{insertprop}=[fill=teal!60]
\tikzstyle{insertget}=[fill=olive!30]
\tikzstyle{insertset}=[fill=magenta!30]
\tikzstyle{insertparam}=[fill=orange!20]
\tikzstyle{insertassign}=[fill=brown!80]
\tikzstyle{updatename}=[fill=green!15]

\pgfdeclarelayer{background}
\pgfsetlayers{background,main}

  \node[template, label={[name=lpre1]\small pre}] (pre1) {};
  \node[template, right of=pre1, xshift=0.5cm, label={[name=lpost1]\small post}] (post1) {};
  \begin{pgfonlayer}{background}
  \node[fit=(pre1)(post1)(lpre1)(lpost1), inner sep=0.2cm, insertprop] (addprop) {};
  \end{pgfonlayer}

  \node[template, right of=pre1, xshift=2.5cm, label={[name=lpre2]\small pre}] (pre2) {};
  \node[template, right of=pre2, xshift=0.5cm, label={[name=lpost2]\small post}] (post2) {};
  \begin{pgfonlayer}{background}
  \node[fit=(pre2)(post2)(lpre2)(lpost2), inner sep=0.2cm, insertparam] (addparam) {};
  \end{pgfonlayer}

  \node[template, right of=pre2, xshift=2.5cm, label={[name=lpre3]\small pre}] (pre3) {};
  \node[template, right of=pre3, xshift=0.5cm, label={[name=lpost3]\small post}] (post3) {};
  \begin{pgfonlayer}{background}
  \node[fit=(pre3)(post3)(lpre3)(lpost3), inner sep=0.2cm, insertassign] (addassign) {};
  \end{pgfonlayer}

  \draw[->] (pre1) -- (post1);
  \draw[->] (pre2) -- (post2);
  \draw[->] (pre3) -- (post3);
  
  \draw[->] (addprop) -- (addparam);
  \draw[->] (addparam) -- (addassign);
  
  \node[right of=post3, xshift=1.5cm, yshift=0.23cm] (spec) {{\scriptsize
    \tabular{l}
    Specification: $\{$\\
    $\quad\version_0 \to \version_3 \to \version_4 \to \version_5$,\\
    $\quad\version_6 \to \version_8 \to \version_9 \to \version_{10}$,\\
    $\quad\version_{11} \to \version_{14} \to \version_{15} \to \version_{16}$\\
    $\}$\\
    \endtabular
  }};
\end{tikzpicture}

\caption{Sketch for the edit sequence pattern ``Insert Property'' $\to$ ``Insert
Parameter'' $\to$ ``Insert Assignment''}
\label{fig:edit-pattern-sketch}
\end{subfigure}

  \caption{\technique: From Edit Graphs to Edit Sequence Patterns. We omit edits $\version_0\to \version_2$ and $\version_{11} \to \version_{13}$ that should be in the edit graphs for ease of presentation.}
  \vspace{-3ex}
\end{figure}


The paths in the quotient graph represent recurrent edit sequences applied by
developers.
For each path in the quotient graph, the \emph{support} is the set of all
edit sequences that correspond to it.
For each path up to a size $n$ with sufficient support in the quotient graph,
\technique creates a sketch along with a \emph{specification} that is given by
its support.
%
%
The right part of Figure~\ref{fig:edit-pattern-sketch} shows the sketch of the
ESP \emph{insert property}, \emph{insert parameter}, and
\emph{assign property}, and the specification given by $\{ \editSeq_1,
\editSeq_2, \editSeq_3 \}$ which correspond to the traces from
Figures~\ref{fig:dev-session},~\ref{fig:dev-session2}, and~\ref{fig:dev-session3}.

\noindent\textbf{From Sketches to Edit Sequence Patterns.}
In the next step, \technique uses these concrete sequences to infer edit
templates and hole predicates to complete the sketch.
We use a procedure based on anti-unification to generalize the edit
sequences into edit templates and corresponding hole predicates.
In essence, anti-unification is a technique to generalize two ASTs into a
template by replacing differing subtrees with holes.
However, we anti-unify edit sequences instead of ASTs and further, generate
predicates relating the holes in the individual templates (see
Section~\ref{sec:antiunify}).

Anti-unifying the edit sequences $\editSeq_1$, $\editSeq_2$, and $\editSeq_3$
produces the ESP depicted in Figure~\ref{fig:pattern}, but without
the predicates $\hole_4' = \hole_4$, $\hole_5' = \mathsf{ToLower}(\hole_5)$, and
$\hole_5^* = \mathsf{ToLower}(\hole_5)$.
This pattern, while general, cannot be used to predict the next changes.
The absence of these predicates means that we can predict neither the name and
type of the inserted parameter, nor the right-hand side of the
assignment.
On the contrary, anti-unifying just $\editSeq_1$ and $\editSeq_2$ produces
exactly the pattern in Figure~\ref{fig:pattern}, which can be used for predictions
as shown in Section~\ref{sec:using-patterns}.
\technique uses agglomerative hierarchical clustering over edit sequences to
produce a hierarchy of increasingly general ESPs.
Hence, we will produce both ESPs (with and without
anti-unifying $\editSeq_3$).
We select and rank a subset of the generated ESPs based
on their predictive power on the input traces.

\ignore{
Note, however, that although $s_1$, $s_2$, and $s_3$ are performing the same high-level operations, $s_3$ differs from $s_1$ and $s_2$ in terms of how the added parameter relates to the added property. Anti-unifying these patterns will lead to a pattern that is too general. 
Although they share the same edit template, there is no predicates for the holes of the name of the added parameter and the right hand side of the added assignment that satisfy all the sequences. 
Without these predicates, it is not possible to use the pattern to predict because we are unable to fill these holes. \mohammad{this discussion is getting a bit abstract - can we possibly show the edit pattern obtained if we simply anti-unify and then show where exactly the problem lies?} \mohammad{Also, I guess the problem is not about it being "too general", but more about it not being useful for prediction since it has free variables towards the end of the pattern? } 
\technique uses hierarchical agglomerative clustering~\cite{getafix} to produce a dendrogram with edit templates and predicates with different levels of generalization. Each sequence of edit templates and its corresponding predicates are used to complete the sketch and create one ESP. In the final step, \technique ranks and filters the ESPs.
}
\section{Edit Sequence Patterns}
\label{sec:prelims}

The goal of this paper is to learn a \emph{sequence} of edit patterns from a set
of developer edit traces and to use the learned patterns to make editing
suggestions while a developer is working in an IDE.
In contrast to related works~\cite{getafix, revisar, yin2019learning} that learn
only a single edit pattern, we aim to leverage the hole predicates among
the sequence of edit patterns. 
In this section, we show a novel representation for the sequence of edit patterns
learned by our approach.

\noindent\textbf{Versions and Development Sessions.}
A \emph{version} $\version$ is a syntactically correct source file that occurs
while a developer is editing code.
A \emph{development session} or \emph{trace} $\Session =
\version_0\ldots\version_n$ is the sequence of all versions that appear during
an editing session.
Here, we identify each version with its \emph{abstract syntax tree} (AST).
Hence, unparsable intermediate versions of code do not appear in the trace.

\noindent\textbf{Edits and Edit Sequences.}
The edit $\edit = \vpre \to \vpost$ changes version $\vpre$ to $\vpost$.
The function $\localize$ on edits that produce the smallest difference between
the two ASTs in the edit.
Formally, $\localize(\vpre \to \vpost) = \vpre^* \to \vpost^*$ if:
\begin{inparaenum}[(a)]
\item $\vpre^*$ and $\vpost^*$ are subtrees of $\vpre$ and $\vpost$,
  respectively;
\item replacing $\vpre^*$ by $\vpost^*$ in $\vpre$ yields $\vpost$; and
\item $\vpre^*$ is the smallest subtree of such kind.
\end{inparaenum}

\begin{example}
  Consider the edit $\version_3 \to \version_4$ in  Figure~\ref{fig:dev-session}, 
  where the developer adds the parameter \code{str id} to the constructor of
  \code{Node}.
  The localized version of this edit $\localize(\version_3 \to \version_4)$ is
  given by $\version_3^* \to \version_4^*$ where:
  \begin{inparaenum}[(a)]
    \item $\version_3^*$ corresponds to the subtree of $\version_3$ that
      represents the empty parameter list \code{()}, and 
    \item $\version_4^*$ corresponds to the subtree of $\version_4$ that
      represents the parameter list \code{(str id)}.
  \end{inparaenum}
  \qed
\end{example}

An edit in the trace $\Session = \version_0\ldots\version_n$ is given by
$\version_i \to \version_j \in \Edits(\Session)$ where $0 \leq i < j \leq n$.
Given edits $\edit = \version_i \to \version_j$ and $\edit' = \version_k \to
\version_\ell$ from $\Session$, we say that $\edit'$ \emph{sequentially follows}
$\edit$ if $i < j = k < l$.
This is written as $\edit \to_\seq \edit'$.
An \emph{edit sequence} $\edit_0\ldots\edit_n$ is a sequence of
contiguous edits,  i.e., $\forall i. \edit_i \to_\seq \edit_{i+1}$.

\noindent\textbf{Templates and Edit Templates.}
An \emph{AST template} (or \emph{template} for short) $\tplt$
is an AST where
some leaf nodes are \emph{holes}, i.e., they do not represent a program
fragment but are placeholders.
A \emph{substitution} $\substitution$ is a function that maps each hole to a
finite sequence of AST nodes.
The AST obtained by replacing each hole $\hole$ in $\tplt$ by the
$\substitution(\hole)$ is written as $\substitution(\tplt)$.
We assume that holes are unique, i.e., that a single hole does not appear in
more than one location in a template and that multiple templates cannot share
holes.

\begin{example}
  \label{ex:ast_templates}
  An example of a template $\tplt$ is \code{($\hole$, str id)} where $\hole$ is
  a hole.
  This template represents all parameter lists of length $1$ or more where the
  last parameter is \code{str id}.
  Note that we are writing templates using the equivalent code for readability.
  The template $\tplt$ is represented as an AST and does not contain a node for
  the comma separator.

  With the substitution $\substitution_0 = \{ \hole \mapsto \epsilon \}$ that
  maps $\hole$ to the empty sequence of nodes, we have $\substitution_0(\tplt)
  = \code{(str id)}$.
  Note that the comma disappears when we substitute the hole with the empty
  list---this is an artifact of writing the template as code.
  For $\substitution_1 = \{ \hole \mapsto \code{int count} \}$ and
  $\substitution_2 = \{ \hole \mapsto \code{int count, str attr} \}$ we have
  $\substitution_1(\tplt) = \code{(int count, str id)}$ and 
  $\substitution_2(\tplt) = \code{(int count, str attr, str id)}$.
  Note that $\substitution_0$, $\substitution_1$, and $\substitution_2$ map
  $\hole$ to sequences of AST nodes of length $0$, $1$, and $2$, respectively.
  \qed
\end{example}

We represent common editing motifs using \emph{edit templates}.
Formally, an edit template $\etplt = \tpltpre \to \tpltpost$ is a pair of AST
templates.
We say that an edit $\vpre \to \vpost$ \emph{matches} an edit template
$\tpltpre \to \tpltpost$ if:
\begin{inparaenum}[(a)]
  \item $\localize(\vpre \to \vpost) = \vpre^* \to \vpost^*$, and
  \item there exists a substitution $\substitution$ such that
    $\vpre^* = \substitution(\tpltpre)$ and $\vpost^* =
    \substitution(\tpltpost)$.
\end{inparaenum}

\begin{example}
  \label{ex:edit-template}
  An example of an edit template is $\etplt = (\hole_1) \to (\hole_2\code{, str id})$.
  Here, the first template matches all parameter lists while the second matches
  all parameter lists where the last parameter is \code{str id}.
  Hence, it would match the edit $\version_3 \to \version_4$ in Figure~\ref{fig:dev-session}.
  However, note that this edit template does not relate the values of $\hole_1$ and
  $\hole_2$ in pre- and post-versions of the edit.
  Therefore, an edit like $\code{(int id)} \to \code{(str label, str id)}$ will
  match the edit template $\etplt$.
  We solve this issue using hole predicates below.
\end{example}

\noindent\textbf{Hole Predicates.}
We introduce the notion of \emph{hole predicates} to
\begin{inparaenum}[(a)]
  \item relate the values of holes across multiple templates, and
  \item restrict the set of substitutions that can be applied to a template.
\end{inparaenum}
Formally, a hole predicate is an expression of type Boolean over holes and
is evaluated over a substitution $\substitution$.
%
%
\begin{compactitem}
  \item \textbf{Unary predicates.} The predicate $\mathsf{IsNotNull}(\hole)$ represents if a hole can be
    replaced by an empty sequence, i.e., the substitution $\substitution$ must satisfy $\substitution(\hole) \neq \epsilon$ if $\mathsf{IsNotNull}(\hole) =
    \mathsf{True}$.
    Another unary predicate $\mathsf{IsKind}_{\mathsf{label}}(\hole)$ is parametrized
    by an AST node type $\mathsf{label}$ (e.g., $\mathsf{AssignExpr}$
    or $\mathsf{ClassDeclaration}$).
    We have that $\mathsf{IsKind}_{\mathsf{label}}(\hole) = \mathsf{True}$ for a
    substitution $\substitution$ only if $\substitution(\hole) =
    \ASTnode$ and the label of $\ASTnode$ is $\mathsf{label}$.
    Note that $\mathsf{IsKind}_{\mathsf{kind}}$ forbids the hole value from
    being an empty sequence and a sequence with multiple elements.
  \item \textbf{Binary predicates.} We also use a class of predicates over two holes, written as
    $\hole_1 = \dependency(\hole_2)$ where $\dependency$ is a function.
    The most common $\dependency$ is the identity function in terms of text value, in which case, we
    write the predicate as $\hole_1 = \hole_2$.
    Other two functions $\dependency$ we use are $\mathsf{ToLower}$ and $\mathsf{ToUpper}$,
    which indicate that the text value of $\hole_1$ in the substitution is the same as
    that of $\hole_2$, but the case of the first character changed appropriately.
\end{compactitem}

\begin{example}[Hole predicates]
  \label{ex:temporal-dependency}
  Consider the template $\tplt = \code{($\hole$, str id)}$ from
  Example~\ref{ex:ast_templates}.
  Here, imposing the predicate $\mathsf{IsNotNull}(\hole)$ 
  ensures that any AST matched by $\tplt$ must have at least $2$
  parameters in the parameter list.
  %
  %
  Continuing from Example~\ref{ex:edit-template}, we can augment the edit
  template $(\hole_1) \to (\hole_2\code{, str id})$ with the hole predicate
  $\hole_1 = \hole_2$ to ensure that we exactly capture the class of edits that
  insert a new parameter \code{str id} to an existing parameter list.
  \qed
\end{example}

\begin{example}
  \label{ex:edit-pattern}
  Hole predicates can be used to relate holes across multiple edits to
  exactly capture the common editing pattern illustrated in Figure~\ref{fig:dev-session}.
  \begin{compactitem}
    \item \emph{Add a new property to a class.}
      This category of edits is captured by the edit template
      $\etplt_1 =
          \code{\{} \hole_1\code{ } \hole_2 \code{\}}
          \to
          \code{\{} \hole_3 \code{ public } \hole_4\code{ } \hole_5 \code{ \{get;\} } \hole_6 \code{\}}$.
      Here, $\hole_1$ and $\hole_2$ represent the class members that appear
      before and after the newly inserted property, respectively.
      The type and name of the property are represented by $\hole_4$ and
      $\hole_5$, respectively.
      We can add the unary predicates $\mathsf{IsKind}_\mathsf{Type}(\hole_4)$
      and $\mathsf{IsKind}_\mathsf{Ident}(\hole_{5})$ to ensure that $\hole_4$ and $\hole_{5}$ are a $\mathsf{Type}$ node and an $\mathsf{Identifier}$ node, respectively.
      To ensure that the contents of the class do not change apart from the
      newly inserted property, we need the predicates $\hole_3 = \hole_1$ and
      $\hole_6 = \hole_2$.
    \item \emph{Add a new parameter to the constructor.}
      This edit is captured by the edit template and predicates similar to
      the ones presented in Example~\ref{ex:temporal-dependency}.
      We have $\etplt_2 = (\hole_7) \to (\hole_8,\code{ }\hole_9\code{ } \hole_{10})$
      with the predicate $\hole_8 = \hole_7$ to ensure that the parameter list
      is preserved apart from the new parameter.
      We have the additional predicates $\hole_9 = \hole_4$ and $\hole_{10} =
      \mathsf{ToLower}(\hole_5)$ to ensure that the type and name match that of
      the inserted property.
      Note that the relation between $\hole_5$ and $\hole_{10}$ is not strict
      equality, but involves an additional transformation $\mathsf{ToLower}$ to
      $\hole_5$.
    \item \emph{Assign the new parameter to the new property.}
      These edits add a new assignment statement to the end of the block and are
      captured by $\etplt_3 = \code{\{} \hole_{11} \code{\}} \to \code{\{}\hole_{12}\code{; }\hole_{13}=\hole_{14}\code{;\}}$
      with the predicates $\hole_{12} = \hole_{11}$, $\hole_{13} =
      \hole_5$, and $\hole_{14} = \hole_{10}$.
      However, we omit some unary predicates if the absence does not hinder understanding for ease of presentation in this paper. 
  \end{compactitem}
  Together, the edit templates $\etplt_1$, $\etplt_2$, and $\etplt_3$ along with
  the above predicates fully capture the common editing pattern of adding a new
  property to a class and initializing it in the constructor.
  \qed
\end{example}

\noindent\textbf{Edit Sequence Patterns.}
The main object of study in this paper is an \emph{Edit Sequence Pattern} (ESP).
ESPs are used to capture sequences of common editing actions
like in Example~\ref{ex:edit-pattern}.
Formally, an ESP is a pair $\langle \tpltseq, \tmprel \rangle$
where:
\begin{inparaenum}[(a)]
  \item $\tpltseq$ is a restricted regular expression over edit templates, and
  \item $\tmprel$ is a set of hole predicates.
\end{inparaenum}
Here, the restricted regular expression $\tpltseq$ is of the form
$\etplt_1\ldots\etplt_{n-1}\etplt_n^{[*]}$ where $[*]$ represents an optional
Kleene star.
That is, $\tpltseq$ is a sequence of edit templates where the last template may
have a Kleene star.

\begin{example}
\label{ex:edit-pattern-seq}

  The edit templates and hole predicates from
  Example~\ref{ex:edit-pattern} can be written as an ESP $\langle \tpltseq, \tmprel \rangle$.
  Here,
    $\tpltseq = \etplt_1 \etplt_2 \etplt_3$ and
    $\tmprel = \{ 
      \hole_3 = \hole_1,
      \hole_6 = \hole_2,
      \hole_8 = \hole_7,
      \hole_9 = \hole_4,
      \hole_{10} = \mathsf{ToLower}(\hole_5),
      \hole_{12} = \hole_{11},
      \hole_{13} = \hole_5,
      \hole_{14} = \hole_{10}
    \} \cup \{
      \mathsf{IsKind}_\mathsf{Type}(\hole_4),
      \mathsf{IsKind}_\mathsf{Identifier}(\hole_{5}),
      \ldots
    \}$.
\end{example}

We say that a sequence of edits $\edit_1 \ldots \edit_n$ matches $\langle
\tpltseq, \tmprel \rangle$, where $\tpltseq =
\etplt_1\ldots\etplt_{n-1}\etplt_n$, if there exists a substitution
$\substitution$ such that:
\begin{inparaenum}[(a)]
  \item each $\edit_i$ matches $\etplt_i$ for $1 \leq i \leq n$,
  \item the hole valuations in $\substitution$ satisfy all the predicates in
    $\tmprel$. 
\end{inparaenum}

Extending this definition, we say that a sequence of edits $\edit_1 \ldots
\edit_m$ (with $m \geq n$) matches  $\langle \tpltseq, \tmprel \rangle$, where
$\tpltseq = \etplt_1\ldots\etplt_{n-1}\etplt_n^*$, if each of the sequences
$\edit_1 \ldots \edit_{n-1} \edit_k$ for $n \leq k \leq m$ matches $\langle
\etplt_1\ldots\etplt_{n-1}\etplt_n, \tmprel \rangle$.

\begin{example}

  Consider an ESP $\langle \tpltseq, \tmprel \rangle$, where
  $\tpltseq = \etplt_1 \etplt_2^*$ has a Kleene star, with 
  $\etplt_1 =
  (\hole_1, \hole_2\ \hole_3) \to (\hole_4),$ $\etplt_2 = (\hole_5, \hole_6) \to
  (\hole_7)$, and
  $\tmprel = \{
    \hole_1=\hole_4,
    \hole_5=\hole_7,
    \mathsf{IsKind}_\mathsf{Type}(\hole_2),
    \mathsf{IsKind}_\mathsf{Ident}(\hole_3),$
  $
    \mathsf{IsKind}_\mathsf{Arg}(\hole_6)
  \}$.
  This pattern represents an editing sequence where the developer deletes the
  last parameter in a declaration, and then, deletes the corresponding
  argument in multiple callsites.

  Consider the three edits $\edit_1$, $\edit_2,$ and $\edit_3$ in
  Figure~\ref{fig:del-param-code}.
  We have that $\edit_1\edit_2\edit_3$ matches $\etplt_1\etplt_2^*$.
  To show this, we need to show that both $\edit_1\edit_2$ and $\edit_1\edit_3$
  match the un-starred ESP $\langle
  \etplt_1\etplt_2, \tmprel \rangle$.
  %
  We can see that $\edit_1\edit_2$ matches $\etplt_1\etplt_2$ with the
  substitution $\substitution_1 = \{
    \hole_1 \mapsto \code{Stream s, byte[] bs},
    \hole_2 \mapsto \code{bool},
    \hole_3 \mapsto \code{flush},
    \hole_4 \mapsto \code{Stream s, byte[] bs},
    \hole_5 \mapsto \code{io, bytes},
    \hole_6\mapsto\code{f},
    \hole_7 \mapsto \code{io, bytes}
  \}$.
  Similarly, we can show that $\edit_1\edit_3$ matches $\etplt_1\etplt_2$ with
  the substitution $\substitution_2$, which is the same as $\substitution_1$
  with $\code{bytes}$ replaced by $\code{result}$ for
  $\hole_5$ and $\hole_7$.
  \qed
\end{example}
\begin{wrapfigure}{l}{0.50\textwidth}
\vspace{-5ex}
{\footnotesize

\hspace{5ex}
\begin{lstlisting}[language=diff, 
        frame=single]
class Comms {
    // Edit 1
-   void Write(Stream s, 
-   byte[] bs, bool flush) { }
+   void Write(Stream s,
+   byte[] bs) { }
}
void Main() {
    // Edit 2
-   Comms.Write(io, bytes, f);}
+   Comms.Write(io, bytes);
    // Edit 3
-   Comms.Write(io, result, f);
+   Comms.Write(io, result);
}
    \end{lstlisting}
    }
\vspace{-3ex}
\caption{Delete a Parameter and Delete Arguments}
\vspace{-5ex}
\label{fig:del-param-code}
\end{wrapfigure}

\begin{remark}
  In our implementation, we consider a slightly more general form of edit
  sequence patterns.
  There, we can have ESPs where any edit template (not just the
  last one) may be starred.
  %
  %
  %
  These more general patterns can be formalized in a straightforward way, though
  we do not do so here for ease of presentation.
\end{remark}

\noindent\textbf{Using Edit Sequence Patterns.}
After a edit sequence matches a prefix of an ESP, we can
use the next edit template in the ESP to predict the next change that the developer will make. 
We illustrate an usage of an ESP in the following example and we will further describe the details in Section~\ref{sec:rank_esp}.

\begin{example}[Usage of an ESP]
  Consider the ESP $\langle \etplt_1\etplt_2\etplt_3, \tmprel
  \rangle$ defined in
  Example~\ref{ex:edit-pattern-seq}, and the edit
  sequence $\edit_1 \to_\seq \edit_2$ from Figure~\ref{fig:dev-session}, where
  $\edit_1 = \version_0 \to \version_3, \edit_2 = \version_3 \to \version_4$. 
  We will consider the task of predicting the next edit give that the
  developer has just performed $\edit_1$ and $\edit_2$.
  \begin{compactitem}
    \item First, we find a substitution $\substitution$ such that $\edit_1$
      and $\edit_2$ match $\etplt_1$ and $\etplt_2$, respectively using
      $\substitution$.
      Further, we require that $\substitution$ satisfies each predicate in
      $\tmprel$ that is over only the holes appearing in $\etplt_1$ and $\etplt_2$.
      Here, we have $\substitution=\{
        \hole_2 \mapsto \code{Node() \{ \}},
        \hole_4 \mapsto \code{str},
        \hole_5 \mapsto \code{Id},
        \hole_6 \mapsto \code{Node() \{ \}},
        \hole_9 \mapsto \code{str},
        \hole_{10} \mapsto \code{id},
      \} \cup \{
        \hole_i \mapsto \epsilon \mid i \in \{1, 3, 7, 8\}
      \}$.
    \item Then, we find an AST node 
    in $\version_4$ such that
      the node matches $\etplt_{3,\pre}$ using a substitution
      $\substitution'$.
      We get $\substitution' = \{ \hole_{11} \mapsto \epsilon \}$ for
      the AST node that represents the empty body of the constructor.
      And we require that $\substitution \cup \substitution'$ satisfies all
      predicates in $\tmprel$ that are over the domain of $\substitution
      \cup \substitution'$.
    \item Now, we use the predicates in $\tmprel$ that contain the holes from
      $\etplt_{3, \post}$ to predict the values for those holes.
      Here, from the predicates $\hole_{12} = \hole_{11}$, $\hole_{13} =
      \hole_5$, and $\hole_{14} = \hole_{10}$, we can predict that $\hole_{12}
      \mapsto \epsilon$, $\hole_{13} \mapsto \code{Id}$, and $\hole_{14} =
      \code{id}$.
    \item Filling in these values in $\etplt_{3, \post}$, we get the new
      constructor body \code{\{Id = id;\}}.
      The predicted version is obtained by replacing $\ASTnode$ in
      $\version_4$ with this new constructor body.
      This exactly produces the version $\version_5$ in Figure~\ref{fig:dev-session}.
  \end{compactitem}
  %
\end{example}

\noindent\textbf{Problem Statement and Solution Sketch.}
The input to the \emph{ESP learning} problem is a set of
traces.
The expected output is a ranked set of ESPs 
$\langle \tpltseq_1, \tmprel_1 \rangle \ldots \langle \tpltseq_n, \tmprel_n
\rangle$.
The aim is to produce ESPs that are helpful in predicting the
next version in any trace.
To this end, we measure the quality of the output using the standard notions of \emph{precision} and \emph{recall}, and a general F score (see Section~\ref{sec:datanalysis} for more details).

\begin{algorithm}
\small
  \begin{algorithmic}[1]
    \Require Set of traces $\Sessions$
    \Ensure Ranked list of ESPs
    \State $\mathsf{edits}
      \gets \bigcup \{ \Edits(\Session) \mid \Session \in \Sessions \}$ \label{line:algo1_step1_start}
    \State $\mathsf{EditGraph} \gets \Call{BuildEditGraph}{\edits}$
    \State $\mathsf{Partitions} \gets$ partition of $\mathsf{edits}$ based on
    $\mathsf{Kind}$ \label{line:algo1_step1_partition_end}
    \State $\mathsf{QuotientGraph} \gets
      \Call{Quotient}{\mathsf{EditGraph}, \mathsf{Partitions}}$ \label{line:algo1_step1_quotient}
    \State $\mathsf{Paths} \gets \Call{FrequentPaths}{\mathsf{QuotientGraph}}$ \label{line:algo1_step1_path_start}
    \State $\mathsf{SketchesAndSpecs}
      \gets \{
        \Call{GenerateSketchAndSpec}{\mathsf{path}}
        \mid
        \mathsf{path} \in \mathsf{Paths}
      \}$\label{line:algo1_step1_path_end}
    \State $\mathsf{Patterns} \gets \emptyset$ \label{line:algo1_step2_start}
    \For{$(\sketch, \spec) \in \mathsf{SketchesAndSpecs}$}
      \State $\mathsf{Patterns} \gets
        \mathsf{Patterns} \cup \Call{LearnPatterns}{\sketch, \spec}$
    \EndFor \label{line:algo1_step2_end}
    \State \Return $\Call{FilterAndSelect}{\mathsf{Patterns}}$ \label{line:algo1_step3}
  \end{algorithmic}
  \caption{Overview of \technique}
  \label{algo:overview}
\end{algorithm}
Our solution strategy is in $3$ parts:
\begin{compactitem}
  \item \emph{Generating edit sequence sketches and specifications.}
    (Section~\ref{sec:generate_sketch}, Lines~\ref{line:algo1_step1_start}-\ref{line:algo1_step1_path_end} in Algorithm~\ref{algo:overview})
    The first step is to generate sets of concrete edit sequences (called the
    specification) that can potentially all match the same ESP, along with a sketch for that ESP.
    %
    %
    To generate these sketches and specifications, we 
    \begin{inparaenum}[(a)]
      \item partition the set of all edits in $\Sessions$ (Lines~\ref{line:algo1_step1_start}-\ref{line:algo1_step1_partition_end}),
      \item summarize the edit graph by the partitions to build a quotient graph (Line~\ref{line:algo1_step1_quotient}), and
      \item generate sketches and specifications from paths of the quotient graph (Lines~\ref{line:algo1_step1_path_start}-\ref{line:algo1_step1_path_end}). 
    \end{inparaenum}
  \item \emph{Synthesizing ESPs.} (Section~\ref{sec:complete_sketch}, Lines~\ref{line:algo1_step2_start}-\ref{line:algo1_step2_end} in Algorithm~\ref{algo:overview})
    Given these edit sequence sketches and specifications, we generate a
    hierarchy of ESPs iteratively where each pattern in the
    hierarchy is more general and matches more edit sequences in the
    specifications than the patterns lower in the hierarchy.
    The core algorithm here takes as input a set of edit sequences and produces
    a set of ESPs that can potentially matches the provided edit sequences.
  \item \emph{Selecting and ranking ESPs.} (Section~\ref{sec:rank_esp}, Line~\ref{line:algo1_step3} in Algorithm~\ref{algo:overview})
    Once we build a hierarchy of ESPs, we determine their
    predictive power by testing them on the input $\Sessions$.
    Based on their precision on the $\Sessions$, we select a subset of
    the patterns and rank them accordingly.
\end{compactitem}

\section{From Traces to Edit Pattern Sketches}
\label{sec:generate_sketch}


In this section, we produce sketches and specifications from a set of traces.
%
Formally, an \emph{edit pattern sketch} $\sketch$ is of the form $A_1\ldots
A_{n-1}A_n^{[*]}$ where each $A_i$ is a placeholder for an edit template.
A \emph{specification} $\spec$ for a sketch $\sketch$ is a
set of edit sequences such that the length of each edit sequence in $\spec$
\begin{inparaenum}[(a)]
  \item is equal to $n$ if $A_n$ is un-starred in $\sketch$, and
  \item is at least $n$ if $A_n$ is starred in $\sketch$.
\end{inparaenum}

\begin{example}
  Given a set of input traces that include the traces from
  Figures~\ref{fig:dev-session} and~\ref{fig:dev-session2}, the technique in this
  section will produce a set of pairs of the form $(\sketch, \spec)$.
  One such pair might be $\sketch = A_1A_2A_3$ and
  $\spec = \{
    \edit_1 \edit_2 \edit_3,
    \edit_1' \edit_2' \edit_3',
    \ldots
  \}$ where
  $\edit_1  = \version_0  \to \version_3$,
  $\edit_2  = \version_3  \to \version_4$,
  $\edit_3  = \version_4  \to \version_5$,
  $\edit_1' = \version_6 \to \version_8$,
  $\edit_2' = \version_8 \to \version_9$, and
  $\edit_3' = \version_9 \to \version_{10}$.
  Note that
  \begin{inparaenum}[(a)]
    \item $\edit_1$ and $\edit_1'$ add a new property,
    \item $\edit_2$ and $\edit_2'$ add a new parameter to the constructor, and
    \item $\edit_3$ and $\edit_3'$ assign the newly added parameter to the newly
      added property.
  \end{inparaenum}
  This sketch and specification will then be used in Section~\ref{sec:complete_sketch} to
  generate a hierarchy of ESPs.
  \qed
\end{example}
We synthesize sketches of ESPs from traces in three steps,
\begin{inparaenum}[(a)]
  \item build an edit graph that contains information about the granularity and
    sequencing of edits in the input traces,
  \item produce a summary of edit sequences by quotienting the edit graph based
    on a partitioning of edits, and
  \item produce sketches and specifications of ESPs by finding frequent paths in
    the summary quotient graph. 
\end{inparaenum}
We explain each of these steps below.

\noindent\textbf{Generating the Edit Graph.}
The edit graph represents all edits in all input traces, as a graph.
First, we collect the set of all edits at all granularities in the input
traces, i.e., edits between all pairs (not necessarily consecutive) of
versions.
Since the number of edits grows quadratically in the length of the
trace, in practice, we prune the edits as follows.
First, we \emph{debounce} the transient edits, i.e., we delete edits where
the two versions were separated by less than $500$ms of time~\cite{Miltner:BluePencil}.
Second, we remove edits where the change is \emph{larger than a given threshold}.
Large edits are likely to incorporate changes that are completely unrelated to each
other.
For example, the edit of adding a new class and implementing all its methods
is likely to contain many unrelated edits, and not be a part of any common
editing workflow.
Now, the individual edits from this pruned set form the vertices of the graph and
there is an edge between $\edit_1$ and $\edit_2$ if and only if $\edit_2$
sequentially follows $\edit_1$, i.e., $\edit_1 \to_\seq \edit_2$.
Note that the edit graph contains edits at different levels of granularity.
For example, in the edit graph for a trace with versions
$\version_0\version_1\version_2$, both the coarse-grained edit $\version_0 \to
\version_2$, as well as the fine-grained edits $\version_0 \to \version_1$ and
$\version_1 \to \version_2$.

\begin{example}
  The edit graph of the trace shown in Figure~\ref{fig:dev-session} contains
  vertices of the form $\graphnode_{ij} = \version_i \to \version_j$ for $0 \leq i < j \leq
  5$.
  The edit graph is shown in Figure~\ref{fig:edit-graph}.
  Note that the graph contains nodes for both fine-grained edits
  $\version_0 \to \version_1$, 
  $\version_1 \to \version_2$, 
  $\version_2 \to \version_3$, and
  $\version_3 \to \version_4$, as well as the coarse-grained edit
  $\version_0 \to \version_4$.
  There is an edge between $\graphnode_{03} \to \graphnode_{34}$ as $\version_3
  \to \version_4$ sequentially follows $\version_0 \to \version_3$.
  On the other hand, there is no edge from $\graphnode_{03}$ to $\graphnode_{45}$.
\end{example}

\noindent\textbf{Summarizing the Edit Graph.}
Once the edit graph is built, the next task is to create an abstract version of
the edit graph that groups together edits of \emph{similar} kind.
To define similar, we first define an embedding of edits.
We categorize edits into $3$ types: insert, delete, and update.
The edit \emph{insert child} $\InsertChild(\parent, \child, i)$ and the edit
\emph{delete child} $\DeleteChild(\parent, \child, i)$ insert and delete AST
node $\child$ of $\parent$'s children at position $i$, respectively, whereas
an \emph{update} $\Update(\old, \new)$ replaces the AST node $\old$ with $\new$.
Insert and delete child operations are also updates (of the parent $\parent$);
however, we assume edits are written as insert or delete child when possible.

Given an edit $\edit$, we define the \emph{kind of the edit}
$\mathsf{Kind}(\edit)$ to be $(\mathsf{operation}, \mathsf{label})$ where:
\begin{inparaenum}[(a)]
  \item $\mathsf{operation}$ is one of $\mathsf{Delete}$,
    $\mathsf{Insert}$, or $\mathsf{Update}$; and 
  \item $\mathsf{label}$ is the type of node that is being deleted, inserted, or
    updated (e.g. $\mathsf{MethodInvocation}$ or $\mathsf{Identifier}$).
\end{inparaenum}
In an edit graph, we call the set of all vertices (edits) of the same kind a
\emph{partition}.

\begin{example}
  \label{ex:edit-graph}
  In Figure~\ref{fig:edit-graph}, the edit $\version_0 \to \version_3$ is of
  type $(\mathsf{Insert}, \mathsf{Property})$ and the type of
  $\version_3 \to \version_4$ is $(\mathsf{Insert}, \mathsf{Parameter})$.
  The partition for $(\mathsf{Insert}, \mathsf{Property})$ is
  given by $\{
    \version_0 \to \version_1,
    \version_0 \to \version_2,
    \version_0 \to \version_3,
    \version_6 \to \version_7,
    \version_6 \to \version_8,
    \version_{11} \to \version_{12},
    \version_{11} \to \version_{13},
    \version_{11} \to \version_{14},
  \}$.
  %
\end{example}

The \emph{quotient graph} of the edit graph summarizes the sequencing information
present in the edit graph at the level of partitions.
We build the quotient graph by lifting the edit graph's sequencing information
to the level of partitions.
\begin{compactitem}
  \item \emph{Quotient graph vertices.}
    A vertex in the quotient graph is a partition, i.e., the set of edits from
    the edit graph with the same $\mathsf{Kind}$.
    We use the term $\Partitions$ to denote the set of all vertices in the
    quotient graph.
  \item \emph{Quotient graph edges.}
    Classically, an edge exists between two vertices $\Partition \to
    \Partition'$ in the quotient graph when there exist $\edit \in \Partition,
    \edit' \in \Partition'$ with an edge $\edit \to_\seq \edit'$ between them
    (see, for example, \cite{bloem2006algorithm}).
    Here, we strengthen the requirement by asking at least $\edseqnumber$
    different pairs of such $\edit$ and $\edit'$.
    This ensures that the ESPs we generate are general, i.e.,
    there are multiple instances of the pattern.
    In our experiments, we use $\edseqnumber = 2$.
  \item \emph{Quotient graph edge labels.}
    We associate each edge $\Partition \to \Partition'$ in the quotient graph
    with a label $\mathsf{Label}(\Partition \to \Partition')$ that is a set of
    edit pairs.
    We define $\mathsf{Label}(\Partition \to \Partition')$ to be
    $\{ (\edit, \edit') \mid \edit \in \Partition, \edit' \in \Partition, \edit
    \to_\seq \edit' \}$, i.e., it contains all pairs of contiguous edits.
\end{compactitem}

\begin{example}
\label{ex:quotient-graph}
  The quotient graph for the edit graph in Figure~\ref{fig:edit-graph} is shown
  in Figure~\ref{fig:quotient-graph}.
  There are $3$ different vertices (partitions) $\Partition_1$, $\Partition_2$,
  and $\Partition_3$ in the quotient graph corresponding to the kinds
  $(\mathsf{Insert}, \mathsf{Property})$,
  $(\mathsf{Insert}, \mathsf{Parameter})$, and
  $(\mathsf{Insert}, \mathsf{Assignment})$, respectively.
  %
  There is an edge $\Partition_1 \to \Partition_2$ from $\Partition_1$ to
  $\Partition_2$ as there are $3 > \edseqnumber$ corresponding sequentially
  consecutive edit pairs:
  \begin{inparaenum}[(a)]
    \item $\version_0 \to \version_3$ and $\version_3 \to \version_4$, 
    \item $\version_6 \to \version_8$ and $\version_8 \to \version_{9}$, and
    \item $\version_{11} \to \version_{14}$ and $\version_{14} \to \version_{15}$. 
  \end{inparaenum}
  Further, the label $\mathsf{Label}(\Partition_1 \to \Partition_2)$ is given
  the same set of $3$ edits.
\end{example}

\noindent\textbf{Generating Sketches and Specifications.}
From the quotient graph, we generate ESP sketches and
corresponding specifications paths using paths in the quotient graph.
First, we define the support $\mathsf{Support}(\Partition_1\ldots\Partition_n)$
as follows:
\begin{inparaenum}[(a)]
  \item For the path with only $1$ edge, we define
    $\mathsf{Support}(\Partition_1\Partition_2)$ to be the label $\mathsf{Label}(\Partition_1 \to \Partition_2$), and
  \item Otherwise, we define
    $\mathsf{Support}(\Partition_1\Partition_2\ldots\Partition_n)$ recursively
    as $\{
      \edit_1 \edit_2 \ldots \edit_n
       \mid
       (\edit_1 \to \edit_2) \in \mathsf{Support}(\Partition_1, \Partition_2),
       (\edit_2 \edit_3 \ldots \edit_n) \in \mathsf{Support}(\Partition_2\ldots \Partition_n)
    \}$.
\end{inparaenum}

Now, we define a \emph{frequent path} in the quotient graph as any path
$\Partition_1\ldots\Partition_n$ where
$\mathsf{Support}(\Partition_1\ldots\Partition_n)$ has cardinality greater than
a threshold $\edseqnumber = \paththreshold$.
The set of frequent paths can be computed recursively by starting with single
edges and adding edges to the end as long as the support is greater than the
threshold.

From the set of frequent paths, we generate two different kinds of sketch
specification pairs.
\begin{compactitem}
  \item For any \emph{simple} path $\Partition_1\ldots\Partition_n$ (i.e.,
    satisfying $i \neq j \implies \Partition_i \neq \Partition_j$), we define 
    $\sketch = A_1 \ldots A_n$ and
    $\spec = \mathsf{Support}(\Partition_1\ldots\Partition_n)$.
  \item For a set of paths
    $\{ \Partition_1\ldots\Partition_{n-1}\Partition_n,
    \Partition_1\ldots\Partition_{n-1}\Partition_n\Partition_n, \ldots,
    \Partition_1\ldots\Partition_{n-1}\Partition_n^k \}$ where 
    $\Partition_1\ldots\Partition_{n-1}$ is simple, we define
    $\sketch = A_1 \ldots A_n^*$ and
    $\spec = \bigcup_{1 \leq i \leq k}
      \mathsf{Support}(\Partition_1\ldots\Partition_{n-1}\Partition_n^i)$.
\end{compactitem}

\begin{example}
  One possible frequent paths in Figure~\ref{fig:quotient-graph} are given by
  $\Partition_1\Partition_2\Partition_3$ where the partitions are equivalent to 
  insert property, insert parameter, and insert assignment as described in
  Example~\ref{ex:quotient-graph}.
  From this path, we generate the sketch $\sketch = A_1A_2A_3$ and the
  specification $\spec = \{
    (\version_0 \to \version_3) (\version_3 \to \version_4) (\version_4 \to \version_5), 
    (\version_6 \to \version_8) (\version_8 \to \version_{9}) (\version_{9} \to \version_{10}), 
    (\version_{11} \to \version_{14}) (\version_{14} \to \version_{15}) (\version_{15} \to \version_{16}) 
  \}$.
\end{example}

Overall, putting together the steps depicted in this section, we
generate a set of sketch-specification pairs $(\sketch, \spec)$ that each
represent a common editing sequence in the input $\Sessions$.

\section{Synthesizing Edit Sequence Patterns}
\label{sec:complete_sketch}

From Section~\ref{sec:generate_sketch}, we get as input a number of 
sketch-specification pairs.
Here, we synthesize a hierarchy of ESPs for each
sketch-specification pair.
The procedure to do this has $3$ major components:
\begin{inparaenum}[(a)]
  \item generate an ESP from an edit sequence,
  \item combine two ESPs to a more general pattern, and 
  \item produce a hierarchy of ESPs using the previous two
    components.
\end{inparaenum}
%
Components (a), (b), and (c) are explained in
Sections~\ref{sec:single-pattern},~\ref{sec:antiunify}, and~\ref{sec:dendrogram},
respectively.

\subsection{Generating an Edit Sequence Pattern}
\label{sec:single-pattern}

Consider generating an ESP from an edit sequence $\edit_1\ldots\edit_n$ and a sketch $\sketch =
A_1\ldots A_n$.
\begin{compactitem}
  \item First, for each edit $\edit_i$, let $\localize(\edit_i) = \edit_{i,\pre}
    \to \edit_{i,\post}$.
    We set $\etplt_i = \edit_{i,\pre} \to \edit_{i,\post}$ and $\tmprel =
    \emptyset$ and then iteratively perform the following operations. 
    \begin{inparaenum}[(a)]
      \item Identify AST nodes $\ASTnode_\pre$ in $\etplt_{i, \pre}$ and
    $\ASTnode_\post$ in $\etplt_{i, \post}$ such that there is a predicate that
    relates the two values.
    Further, we pick $\ASTnode_\pre$ and $\ASTnode_\post$ such that they are of
    the largest possible size.
    For example, we may pick $\ASTnode_\pre$ and $\ASTnode_\post$ such that
    $\ASTnode_\post = \ASTnode_\pre$ or $\ASTnode_\post =
    \mathsf{ToLower}(\ASTnode_\pre)$.
    \item We replace $\ASTnode_\pre$ and $\ASTnode_\post$ in $\etplt_i$ by two
    fresh holes $\hole_\pre$ and $\hole_\post$, and add the predicate that
    relates the two
    values to $\tmprel$ (e.g., $\hole_\post = \hole_\pre$ or $\hole_\post =
    \mathsf{ToLower}(\hole_\pre)$).
    We will also add unary predicates $\mathsf{IsNotNull}$ and $\mathsf{IsKind}$ of $\hole_\post$ and $\hole_\pre$ if they satisfy the constraints. 
    \item We add the generated mappings to a substitution $\substitution_i$.
    \end{inparaenum}
    %
  \item Then, we union all the substitutions $\substitution_i$ into a single
    $\substitution$ (note that domains of $\substitution_i$ are disjoint).
    Now, for each $\hole_1 \to \ASTnode_1, \hole_2 \to \ASTnode_2 \in
    \substitution$, we check if there exists a predicate that relates
    $\ASTnode_1$ and $\ASTnode_2$.
    If so, we add that predicate on $\hole_1$ and $\hole_2$ to the $\tmprel$.
\end{compactitem}
The produced ESP is $\langle \etplt_1\ldots\etplt_n, \tmprel
\rangle$.
Note that the above procedure is for sketches without Kleene stars---we discuss
how to generate patterns with Kleene stars later.

\begin{example}
  \label{ex:leaf_nodes}
  Consider the edit sequence $\edit_1 \edit_2 \edit_3$ where $\edit_1 =
  \version_6 \to \version_8$, $\edit_2 =
  \version_8 \to \version_9$, and $\edit_3 =
  \version_9 \to \version_{10}$ from Figure~\ref{fig:dev-session2}.
  %
  We illustrate the ESP generation procedure for $\edit_1$.
  First, we start with the localized edit $\localize(\edit_1) = 
  \code{\{<IdDecl> <Ctor>\}}
  \to
  \code{\{<IdDecl> public int Size \{get; set;\} <Ctor>\}}
  $.
  The terms $\code{<IdDecl>}$ and $\code{<Ctor>}$ are shorthand for
  $\code{public int Id \{get; set;\}}$ and $\code{Graph(int id) \{Id = id;\}}$,
  respectively.

  Now, over all pairs of nodes in the localized edit, we check if there is a
  predicate that is satisfied by the nodes.
  Here, we get that the node $\code{<IdDecl>}$ is repeated in both the pre- and
  post-versions.
  Replacing these with holes, we get the edit template $
    \code{\{}\hole_1 \code{ <Ctor>\}}
    \to
    \code{\{}\hole_3\code{ public int Size \{get; set;\} <Ctor>\}}
  $ and the predicate $\hole_1 = \hole_3$. 
  Repeating this, we replace $\code{<Ctor>}$ with $\hole_2$ and $\hole_4$
  to get the edit template $\etplt_1 =
    \code{\{}\hole_1\ \hole_2\code{\}}
    \to
    \code{\{}\hole_3 \code{ public int Size \{get; set;\} }\hole_4\code{\}}
  $ and the predicate $\hole_4 = \hole_2$. 

  Doing the similar procedure on $\edit_2$ and $\edit_3$, we get the edit
  templates
  $\etplt_2 = (\hole_5) \to (\hole_6\code{, int size})$ and 
  $\etplt_3 = \code{\{}\hole_7\code{\}} \to \code{\{}\hole_8\code{ Size = size;\}}$, with the predicates 
  $\hole_6 = \hole_5$ and $\hole_8 = \hole_7$.
  We then compute the predicates across the different $\etplt_i$, and in this
  example, we do not find any.

  The ESP returned is $\etplt_1 \etplt_2 \etplt_3$ along with
  the predicates $\tmprel = \{
    \hole_3 = \hole_1,
    \hole_4 = \hole_2,
    \hole_6 = \hole_5,
    \hole_8 = \hole_7
  \}$.
  Note that the pattern does not create holes for the type or name of the
  inserted property or parameter (\code{int}, \code{Size}, and \code{size}).
  With just a single edit sequence, we do not have any evidence for the need to
  generalize these identifiers---hypothetically, every property that is added in
  the input traces might have the type $\code{int}$ and name $\code{Size}$.
  We can generalize these identifiers when we have two ESPs as
  we will describe next.
  \qed
\end{example}

\subsection{Combining Edit Sequence Patterns}
\label{sec:antiunify}

Using the previous step, we can generalize all concrete edit sequences in
$\spec$ to ESPs. 
Now, we discuss how to combine any two such ESPs into a
single, more general, ESP.
The primary tool we use for this purpose is \emph{anti-unification}~\cite{Plotkin}. 
Anti-unification is a classical operation of trees that retains the parts that
are common to two trees, while replacing the parts that are different with
holes.
It has been used in code edit analysis and synthesis literature to generalize
ASTs and edits in multiple contexts~\cite{getafix,onepointone,revisar}.
However, in our technique, we need to anti-unify sequences of edit templates 
rather than
ASTs or single edit templates, and further, need to consider the predicates.

Formally, given two sequences of edit templates $\etplt_1\ldots\etplt_n$ and
$\etplt_1'\ldots\etplt_n'$, the anti-unification of the two sequences produces an
edit template sequence $\etplt_1^*\ldots\etplt_n^*$ and two substitutions
$\substitution, \substitution'$ such that:
For each $\etplt_i^*$, we have that
$\etplt_i = \substitution(\etplt_{i,\pre}^*) \to \substitution(\etplt_{i,\post}^*)$ and 
$\etplt_i' = \substitution'(\etplt_{i,\pre}^*) \to \substitution'(\etplt_{i,\post}^*)$.
Intuitively, anti-unification is generalization: if any edit sequence
$\edit_1\ldots\edit_n$ matches $\etplt_1\ldots\etplt_n$ or
$\etplt_1'\ldots\etplt_n'$, it will also match $\etplt_1^*\ldots\etplt_n^*$.
Further, we also need to generalize hole predicates $\tmprel$ and $\tmprel'$.
For this, we generate only those hole predicates that are satisfied by both substitutions $\substitution$ and $\substitution'$.
As a result, the newly generated hole predicates is also a generalization.

\begin{example}[Anti-unification of edit sequence patterns]
  \label{ex:combine_nodes}
  Recall the edit templates $\etplt_1 \etplt_2 \etplt_3$ from
  Example~\ref{ex:leaf_nodes}.
  Now consider another ESP from a similar sequence of edits shown in Figure~\ref{fig:dev-session}, but
  with the following changes:
  \begin{inparaenum}[(a)]
    \item the property and parameter added has a different name and type (\code{str Id} and \code{str id}) and
    \item the parameter list in the constructor and the body of the constructor are empty.
  \end{inparaenum}
  In the edit templates $\etplt_1' \etplt_2' \etplt_3'$ for an ESP generated for this case, $\etplt_1'$ is similar to
  $\etplt_1$ with the hole names replaced.
  However, $\etplt_2' = \code{()} \to \code{(str id)}$, i.e., it does not
  have $\hole_5$ and $\hole_6$ to represent the already existing parameters in
  the parameter list.
  Similarly, $\etplt_3' = \code{\{ \}} \to \code{\{Id = id;\}}$ and it does not contain $\hole_7$ and $\hole_8$.

  To generalize the two edit templates $\etplt_1 \etplt_2 \etplt_3$ and $\etplt_1'
  \etplt_2' \etplt_3'$, we anti-unify the before and post templates in each $\etplt_i$ and $\etplt_i'$ one by one:
  \begin{compactitem}
    \item For $\etplt_1$ and $\etplt_1'$, we get the anti-unified edit template
      $\etplt_1^* = \code{\{}\hole_1^*\code{ } \hole_2^* \code{\}} \to \code{\{}\hole_3^*\code{ public }\hole_9^*\code{ } \hole_{10}^*\code{ \{get; set;\} }\hole_4^*\code{\}}$.
      Note that the type and name of the properties have been replaced by new
      holes $\hole_9^*$ and $\hole_{10}^*$.
     \item For generalizing, $\etplt_2$ and $\etplt_2'$, additional care must be
      taken as there are no holes corresponding to $\hole_5$ and $\hole_6$ in
      $\etplt_2'$.
      Some anti-unification approaches~\cite{getafix, revisar} will produce an
      overly general edit template $\etplt_2^*$ as $(\hole_{11}^*) \to
      (\hole_{12}^*)$.
      With this edit template, we do not have any holes for the name and type of
      the parameter, and thus we cannot express the hole predicate between parameter type and property type.

      We propose to further generalize the lists of children, i.e., ($\hole_6$\code{, int size}) and (\code{str
      id}), inspired by~\citet{onepointone}.
      During anti-unification, we examine if the two children lists can be
      better generalized by introducing additional holes which are substituted
      by the empty token $\epsilon$ in one case.
      Doing so, we get $\etplt_2^* = (\hole_5^*) \to (\hole_6^*\code{, }\hole_{11}^*\code{ }\hole_{12}^*)$.
    \item Similarly, for $\etplt_3$ and $\etplt_3'$, we get
      $\etplt_3^* = \code{\{}\hole_7^*\code{\}} \to \code{\{}\hole_8^*\code{ }\hole_{13}^* = \hole_{14}^*\code{\}}$.
  \end{compactitem}

 
Note that we can get substitutions $\substitution$ and $\substitution'$ for
free after we generalized these edit templates.
These substitutions can then be used to generate the hole predicates.
  \begin{inparaenum}[(a)]
    \item For a specific unary hole predicate $\dependency$, we enumerate every hole $\hole$ in generalized edit templates and checking whether both $\dependency(\substitution(\hole))$ and $\dependency(\substitution'(\hole))$ are satisfied. 
    Notice that $\substitution(\hole)$ and $\substitution'(\hole)$ can still contain holes in the first or the second ESP. 
    If there are any holes in $\substitution(\hole)$ or $\substitution'(\hole)$, we recursively repeat the substitution procedure until $\hole$ maps to an AST in concrete edit sequences.
    \item We generate binary hole predicates similarly but enumerate all pairs of holes in generalized edit templates.
      In this case, we end up with the starred version of the predicates in
      $\tmprel$ from Example~\ref{ex:leaf_nodes}, along with the predicates $\{
        \hole_{11}^* = \hole_{9}^*,
        \hole_{12}^* = \mathsf{ToLower}(\hole_{10}^*),
        \hole_{13}^* = \hole_{10}^*,
        \hole_{14}^* = \hole_{12}^*
      \}$.
  \end{inparaenum}
  
  The generalized edit templates $\etplt_1^* \etplt_2^* \etplt_3^*$ along with the
  new predicates exactly capture the editing sequence of adding a new property
  and initializing it in the constructor.
  \qed
\end{example}
As illustrated in the previous example, we cannot generalize ESPs using standard anti-unification techniques.
When two nodes in the edit templates have different numbers of children, we may
need to introduce new holes that map to $\epsilon$. 
We do not explicitly write out our anti-unification algorithm here for the lack
of space---instead, it is available in the supplementary material.
The algorithm takes as input two ESPs
$\langle \tpltseq_1,\tmprel_1 \rangle$ and 
$\langle \tpltseq_2,\tmprel_2 \rangle$, and produces a more general
$\langle \tpltseq, \tmprel \rangle$.
Along with the generalized ESP, the algorithm also returns a
cost of anti-unification.
This cost roughly measures how general the $\langle \tpltseq, \tmprel \rangle$
is compared to $\langle \tpltseq_1, \tmprel_1 \rangle$ and $\langle \tpltseq_2,
\tmprel_2 \rangle$, with more general patterns getting a higher cost than less
general ones.
This corresponds to the intuition that anti-unification algorithms attempt to
compute the least general generalization of two objects.
Based on the anti-unification costs, we will generate a hierarchy of ESP in the following section.

%
%

\subsection{Building a Hierarchy of Edit Sequence Patterns}
\label{sec:dendrogram}

\smallskip
\begin{algorithm}
  \small
  \begin{algorithmic}[1]
    \Require Sketch of edit sequence pattern $\sketch$
    \Require Specification for edit sequence pattern $\spec$
    \Ensure A set of edit sequence patterns $\mathsf{Patterns}$
    \State $\mathsf{Nodes} \gets \{
      \Call{GeneratePattern}{\sketch, \edit_1\ldots\edit_n}
      \mid
      \edit_1\ldots\edit_n \in \spec
    \}$ \label{line:algo2_step1}
    \State $\mathsf{Patterns} \gets \varnothing$
    \While{$|\mathsf{Nodes}| > 1$}
      \State Pick $\mathsf{Node}_1, \mathsf{Node}_2$ such that
        $\Call{AntiUnifyCost}{\mathsf{Node}_1, \mathsf{Node}_2}$ is minimal \label{line:algo2_step2_start}
      \State $\mathsf{NewNode} \gets \Call{AntiUnify}{\mathsf{Node_1}, \mathsf{Node_2}}$ \label{line:algo2_step2_end}
      %
      \State $\mathsf{Patterns} \gets \mathsf{Patterns} \cup \{\mathsf{NewNode}\}$
      \State $\mathsf{Nodes} \gets \mathsf{Nodes} - \{\mathsf{Node}_1, \mathsf{Node}_2\} \cup \{\mathsf{NewNode}\}$
    \EndWhile
    \State \Return $\mathsf{Patterns}$
  \end{algorithmic}
  \caption{The procedure of building a dendrogram ($\mathsf{LearnPatterns}$ in Algorithm~\ref{algo:overview})}
  \label{algo:build_dendrogram}
\end{algorithm}

Algorithm~\ref{algo:build_dendrogram} shows the full procedure going from an ESP sketch $\sketch$ and specification $\spec$ to a set of ESPs $\mathsf{Patterns}$.
The algorithm performs a standard agglomerative hierarchical
clustering (AHC)~\cite{AHC} with the distance metric given by the
anti-unification cost.
AHC builds a dendrogram where each node is an ESP.
This is reminiscent of the techniques~\cite{getafix}, but performed over a
sequence of edits rather than a single one.
At line~\ref{line:algo2_step1}, we build the leaf nodes in dendrogram by generalizing edit sequences
to ESPs as described in Section~\ref{sec:single-pattern}.
At lines~\ref{line:algo2_step2_start}-\ref{line:algo2_step2_end}, we select two nodes in the dendrogram $\mathsf{Node_1},
\mathsf{Node_2}$ that have the lowest merging anti-unification cost in
anti-unification and anti-unify them into a new dendrogram node
$\mathsf{NewNode}$.
The procedure eventually returns the set of all nodes that were constructed.


%

\begin{remark}[Handling Kleene stars]
An ESP sketch $A_1\ldots A_{n-1}A_n^{[*]}$ with Kleene stars may contain edit sequences with various lengths.
To handle the sketch with Kleene stars, we will compute a new set of edit sequences, each of which has length $n$ equal to the sketch, such that we can build the dendrogram in the same way as the sketches without Kleene stars.
Concretely, for an edit sequence $\edit^i_1 \ldots \edit^i_m$ in the sketch we will collect all subsequences $\edit^i_1 \ldots \edit^i_{n-1} \edit^i_{k}$ for all $n \le k \le m$.

Suppose we have an ESP sketch $A_1A_2^*$ and an edit sequence $\edit_1\edit_2\edit_3\edit_4$, where $\edit_1$ corresponds to $A_1$ and $\edit_2, \edit_3, \edit_4$ correspond to $A_2$. 
We will break the edit sequence into three edit sequence with length $2$, namely, $\edit_1\edit_2$, $\edit_1\edit_3$, and $\edit_1\edit_4$.
\end{remark}

\section{Ranking Edit Sequence Patterns}
\label{sec:rank_esp}

As we discussed in Section~\ref{sec:motivating-example}, we cannot simply pick more
general ESPs over less general ones.
More general patterns may be less predictive than specific ones.
In this section, we will select a ranked list of ESPs from all
dendrograms as the output of \technique.


\begin{algorithm}
  \small
  \begin{algorithmic}[1]
    \Require An edit sequence pattern
        $\langle \tpltseq=\etplt_1\ldots\etplt_{n-1}\etplt_n^{[*]} , \tmprel \rangle$
    \Require A trace $\version_0\ldots\version_m$
    \Ensure Prediction for next version $\hat{\version}$ or $\bot$
    
    
    \ForAll{possible edit sequences $\edit_1\to_\seq\ldots\to_\seq\edit_k$ ending at $\version_m$ and $k\le n$}\label{line:algo3_match_start}  
        \If{$\edit_1\ldots\edit_k$ matches $\etplt_1\ldots\etplt_k$ using a \emph{unique} $\substitution$} \label{line:algo3_match_end}  
            \If{$k < n$}\label{line:algo3_pred_start}  
                \State $\mathsf{v_\mathsf{predicted}}
                    \gets \Call{Predict}{\etplt_{k+1}, \substitution}$
                \If {$\mathsf{v_\mathsf{predicted}} \neq \bot$}
                    \Return $\mathsf{v_\mathsf{predicted}}$
                \EndIf
            \Else
                \State $\mathsf{v_\mathsf{predicted}}
                    \gets \Call{Predict}{\etplt_{n}, \substitution}$
                \If {$\mathsf{v_\mathsf{predicted}} \neq \bot$}
                    \Return $\mathsf{v_\mathsf{predicted}}$
                \EndIf
            \EndIf \label{line:algo3_pred_end}  
        \EndIf
    \EndFor
    \State \Return $\bot$

    \Function{Predict}{$\etplt = \tpltpre \to \tpltpost$, $\substitution$}
        \ForAll{Every subtree $\version_m^*$ in $\version_m$} 
            \If{$\exists \substitution'. \substitution'(\tpltpre)=\version_m^* \wedge \substitution' \text{ satisfy } \tmprel \wedge \substitution \subseteq \substitution'$}
                \State \Return $\version_m$ with the subtree $\version_m^*$ replaced by $\substitution'(\tpltpost)$
            \EndIf
        \EndFor
        \State \Return $\bot$
    \EndFunction

  \end{algorithmic}
  \caption{The procedure of edit sequence pattern prediction}
  \label{algo:use-edit-pattern}
\end{algorithm}

\noindent\textbf{Predictions Using Edit Sequence Patterns}
%
%
First, we discuss how an ESP can be used for predicting the
next change in Algorithm~\ref{algo:use-edit-pattern}.
As input, it takes an ESP $\langle \tpltseq, \tmprel \rangle$
and a trace $\version_0\ldots\version_m$.
Algorithm~\ref{algo:use-edit-pattern} will first match a developer's edits
against a prefix of a learned ESP (Lines~\ref{line:algo3_match_start}-\ref{line:algo3_match_end}) and use the next
edit template in the ESP to predict the change the developer
is going to make next (Lines~\ref{line:algo3_pred_start}-\ref{line:algo3_pred_end}).

\begin{remark}
In our implementation, instead of brute-force enumeration at Lines~\ref{line:algo3_match_start}-\ref{line:algo3_match_end}, we enumerate all matched edit sequences by prefixes of an
ESP in polynomial time.
We do this by matching edits on a deterministic
finite automaton generated from the ESP. 
Also note that we require the substitution $\substitution$ to be \emph{unique}
to avoid over-generalization. 
In our evaluation, we provide with the cursor location information to
\code{Predict} in Algorithm~\ref{algo:use-edit-pattern} such that we only need
to enumerate subtrees $\version_m^*$ that contain the cursor location of the
user to make more precise predictions. 
Notice that we only let an ESP generate its first prediction for a trace in Algorithm~\ref{algo:use-edit-pattern}. However, we recorded all predictions that can be made by an ESP in our evaluation and found that no ESP had made multiple predictions for a trace.
\end{remark}
For each ESP, there are three outcomes at each version $\version_m$: \begin{inparaenum}
\label{sec:prediction}
  \item the pattern does not predict, i.e., returns $\bot$,
  \item \emph{correct} prediction, the pattern predicts $\hat{\version}$ that is
    equal to $\version_{l},\ l>m$ in the later sessions, and
  \item otherwise, we consider the pattern makes a \emph{wrong} prediction.
\end{inparaenum}

\noindent\textbf{Ranking and Selecting Edit Sequence Patterns.}
From Section~\ref{sec:complete_sketch}, we get a set of ESPs
$\mathsf{Patterns}$.
%
Let $\editSequences$ be the union of all edit sequences in the specifications generated in Section~\ref{sec:generate_sketch}.
Here, we try to solve the following problem: select and rank a subset of $\mathsf{Patterns}$
such that we maximize the correctly predicted versions and minimize the wrongly
predicted versions.
Algorithm~\ref{algo:rank-pattern} depicts the procedure for this.

\begin{algorithm}
  \small
  \begin{algorithmic}[1]
    \Require A list of edit sequence patterns $\mathsf{Patterns}$
    \Require Input traces $\Sessions$
    \Require Edit sequences $\mathsf{EditSeqs}$
    \Require Precision thresholds $\mathsf{threshold}_1, \mathsf{threshold}_2$
    \Ensure A ranked list of edit sequence patterns $\mathsf{Patterns}$

    \Function{FilterAndSelect}{}
    \State $\mathsf{Patterns} \gets$ \Call{GreedySelect}{$\mathsf{Patterns}$, $\mathsf{EditSeqs}$, $\mathsf{threshold}_1$} \label{line:algo4_select_edit}
    \State $\mathsf{Patterns} \gets$ \Call{GreedySelect}{$\mathsf{Patterns}$, $\Sessions$, $\mathsf{threshold}_2$} \label{line:algo4_select_session}
    \State \Return $\mathsf{Patterns}$
    \EndFunction

    \Function{GreedySelect}{$\mathsf{Patterns}$, $\mathsf{Data}$, $\mathsf{threshold}$}
    \State $\mathsf{SelectPatterns} \gets [], \mathsf{Uncovered} \gets \mathsf{Data}$
    \ForAll {$p_i \in \mathsf{Patterns}$}
      \State $\mathsf{correct_i}, \mathsf{incorrect_i} \gets
        \Call{Evaluate}{p_i, \mathsf{Data}}$
    \EndFor
    \State $\mathsf{Patterns} \gets \{
      p_i \mid \frac{\mathsf{correct_i}}{\mathsf{correct_i} + \mathsf{incorrect_i}}> \mathsf{threshold}
    \}$
    \While{$\mathsf{Patterns} \neq \varnothing \wedge \mathsf{Uncovered} \neq \varnothing$}
        \State $p
          \gets \mathsf{argmax}_{p_i \in \mathsf{Patterns}} \mathsf{correct_i} - \mathsf{incorrect_i}$ 
        \State $\mathsf{Patterns} \gets \mathsf{Patterns} - \{p\}$
        \State $\mathsf{SelectPatterns} \gets \mathsf{SelectPatterns} + [p]$
        \State $\mathsf{Uncovered} \gets \mathsf{Uncovered}
                - \{\text{data points covered by }p\}$
        \State Update all $\mathsf{correct_i}$ and $\mathsf{incorrect_i}$ according to $\mathsf{Uncovered}$
    \EndWhile
    \State \Return $\mathsf{SelectPatterns}$
    \EndFunction
  \end{algorithmic}
  \caption{The procedure of ranking edit sequence patterns ($\mathsf{FilterAndSelect}$ in Algorithm~\ref{algo:overview})}
  \label{algo:rank-pattern}
\end{algorithm}

The core component of Algorithm~\ref{algo:rank-pattern} is the
$\mathsf{GreedySelect}$ procedure.
The procedure takes as input a set of patterns $\mathsf{Patterns}$ and traces
$\mathsf{Data}$ and produces a ranked subset of patterns based on
their predictive performance.
Intuitively, $\mathsf{GreedySelect}$ works similar to the approximate set-cover
algorithm.
We call each version in a trace in $\mathsf{Data}$ a data point.
The procedure maintains a partial list of selected patterns and a set
$\mathsf{Uncovered}$ of data points on which no selected pattern has made a prediction.
We first measure the number of correct and incorrect
predictions each pattern makes on $\mathsf{Data}$.
With these count of correct and incorrect predictions, we then eliminate all
patterns with precision less than a given threshold.
In each iteration of selection,
\begin{inparaenum}[(a)]
  \item we update the partial list of pattern with the best pattern as measured
    by the difference in the number of correct and incorrect predictions, and
  \item we remove the set of datapoints on which the best pattern made predictions from $\mathsf{Uncovered}$ and update all $\mathsf{correct_i}$ and $\mathsf{incorrect_i}$ accordingly.
\end{inparaenum}

Algorithm~\ref{algo:rank-pattern} calls $\mathsf{GreedySelect}$ twice in two
phases.
In the first, we only consider the predictive power of each pattern on the set
$\editSequences$.
Then, in the second step, we select and rank based on the full training data,
i.e., the input $\Sessions$.
Ideally, we only need the second step at line~\ref{line:algo4_select_session} because the precision on traces reflects the effectiveness of ESPs in real scenarios.
However, we add the first step of filtering to reduce the number of patterns evaluated in the second step because performing $\mathsf{GreedySelect}$ over traces on all patterns is very expensive.
We find the first step is able to filter out most of the patterns, speeding up the second step by a large degree.


\section{Evaluation}

In this section, we present our evaluation to address the following research questions:
\begin{itemize}

    \item \textbf{RQ1}: How effective is \technique at predicting edit sequences performed in the IDE?
    
    \item \textbf{RQ2}: What kind of ESPs are learned by \technique?
\end{itemize}

\subsection{Experimental Setup}
\label{sec:datanalysis}


\noindent\textbf{Data Collection.}
We developed a Visual Studio extension to record all syntactically correct versions of
the documents updated by the developer (in the background, without interruption).
We selected Visual Studio as the target IDE in this study as it is the most
popular IDE for C\#.
\ignore{
Following an approach proposed by Miltner et al.~\cite{Miltner:BluePencil}, we reduced the number of document versions by discarding versions that occur within 500ms of each other. This process avoids mining edits that are extremely fine-grained and likely \emph{transient}, such as when the developer is typing the name of a variable character-by-character.
}
We contacted \textbf{\numberOfDevelopers} professional software developers from
a large software company who agreed to use the extension and participate in the
study. 
They were working on \textbf{four} separate C\# code bases with a total of $\mathsf{377.5K}$
source lines of code.
Initially, we recorded $\mathbf{\noOfsessions}$ \emph{development sessions}
($\cong$ $\mathsf{250}$ hours) containing $\mathbf{\totalVersions}$ \textit{versions}
of $\mathbf{\noOfDocuments}$ \textit{documents}, which we refer to as
\emph{training dataset}.
After $\mathsf{6}$ months, we collected an additional $\mathbf{399}$ sessions (containing
$\mathbf{201,142}$ versions), which we refer to as \emph{test dataset}. 

\ignore{Since the $\mathsf{SelectRankBySession}$ procedure is a more expensive way to select patterns, we chose different thresholds for this procedure to evaluate its benefits. }

\noindent\textbf{Parameter Choices.}
Apart from data, \technique requires a few runtime parameters :
\begin{inparaenum}[(a)]
\item The maximum length of the sequences, 
\item minimum length of the edit sequences (Section~\ref{sec:generate_sketch}),
\item thresholds $\mathsf{threshold}_1$ and $\mathsf{threshold}_2$
    for selecting patterns based on the sketch and session analysis,
    respectively (Section~\ref{sec:complete_sketch}).
\end{inparaenum}
We chose \texttt{n = 3} for the maximum length based on our initial set of sequence examples,
which did not contain longer sequences, and \texttt{\edseqnumber{} = 2} for the support so
that we have at least two examples for each pattern.
We study the effect of varying the values of $\mathsf{threshold}_1$ and
$\mathsf{threshold}_2$ in the experiments for RQ1.

\noindent\textbf{RQ1 Methodology.}
To answer RQ1, we
\begin{inparaenum}[(a)]
  \item perform a 5-fold validation over our \emph{training dataset} and
  \item  simulate the learned patterns from the training dataset on
    our \textit{test dataset} containing unseen development sessions.
\end{inparaenum}  
To perform the $\mathsf{5}$-fold validation, we randomly split \textbf{\noOfsessions} training
sessions into $\mathsf{5}$ equal \emph{folds} and
for each fold, we evaluate the ESPs that were learned from the other $\mathsf{4}$ folds.
We repeat the $\mathsf{5}$-fold validation varying $\mathsf{threshold}_1, \mathsf{threshold}_2
\in \mathsf{[0,1]}$ by steps of $\mathsf{0.1}$.
Using the best threshold values found (as measured by the $F_3$ metric
defined below), we evaluate the ESPs learned from the full training dataset on
the test dataset.

For each experiment, we compute the \textit{precision} of the code edit suggestions as
the proportion of the total suggestions that are \textit{correct}~(see \Cref{sec:prediction}).
To compute \textit{recall}, we need an oracle containing all the edits expected
from \technique.
Since our data set consists of thousands of fine-grained edits, it is not feasible
to construct such an oracle manually.
Instead for our experiments, we define our baseline as the number of correct
suggestions we get when using \technique in its most general setting with
$\mathsf{threshold}_1 = 0$ and $\mathsf{threshold}_2 = 0$ values in
Algorithm~\ref{algo:rank-pattern}.
Using this baseline, we define relative recall (\textit{recall}$_{rel}$) as the
ratio of correct prediction \technique produces at a given configuration with
respect to the baseline. 

To consolidate the \textit{precision} and \textit{recall} metric into a single score, we make use of a variation of the popular $F_{\beta}$
%
metric \cite{Chinchor1992MUC4,InformationRetrievalBlair}.
Here, we use $F_{\gamma}$ given by:
\begin{equation}
    F_{\gamma} = \frac{(1+\gamma^2)*precision*recall_{rel}}{ precision + \gamma^2 recall_{rel}}
\end{equation}
Note that this definition is equivalent to the definition of $F_\beta$ with
$\beta$ set to $\frac{1}{\gamma}$ and the recall term replaced by  relative recall.
The $\gamma$ parameter allows us to choose the relative emphasis we put on the  \textit{precision} term compared to \textit{recall}  term, with \textit{precision} given $\gamma$ times more importance over \textit{recall}~\cite{InformationRetrievalBlair}. 
For our evaluation, we make use of $F_{\gamma = 3}$ to give \textit{precision} 3 times more importance than \textit{recall}$_{rel}$.
We chose to increase the emphasis on precision because of two reasons: 
\begin{inparaenum}[(a)]
    \item Reliability is one of main causes of disuse of automated refactorings in IDEs~\cite{6227190}, so tool builders tend to favor precision over recall.  
    \item The use of \textit{recall}$_{rel}$ instead of \textit{recall} tend to introduce a bias towards recall because \textit{recall}$_{rel}$ will be higher than ground-truth \textit{recall}.
\end{inparaenum}

For notational simplicity we use $F_{3}$ instead of $F_{\gamma = 3}$ to refer to this metric in the rest of the paper.

%



\noindent\textbf{RQ2 Methodology.}
\label{sec:RQ2Setup}
To answer RQ2, we manually analyzed the ESPs learned by \technique using the best threshold
parameters found from the RQ1 study.
Two authors, each with more than 5 years of professional development in C\#, coded these
ESPs using established guidelines from the literature~\cite{Saldana:coding, Campbell:2013:CodingSemiStructured}.
They first iteratively refined the code set on 20\% of the patterns.
Then, using this code set, they independently coded another 20\% of the patterns.
We then use Cohen's kappa to calculate inter-rater reliability.
Their inter-rater reliability was 0.95, which shows a high agreement between the two raters.
We then split the rest of the patterns into two sets, and they independently coded each set.
We detail each code set in RQ2.

\subsection{Results}
\label{sec:results}

\noindent\textbf{\emph{RQ1: How effective is \technique at predicting edit sequences performed in the IDE?}}
\vspace{0.1cm}

\begin{table}[t]
\centering
\small
\caption{Precision and relative recall of \technique sorted by 5-fold Validation $F_3$ scores}
\label{tab:precision}
\vspace{-2ex}
\begin{tabular}{cc|ccc|ccc}
\multirow{3}{*}{Threshold$_1$ } &
\multirow{3}{*}{Threshold$_2$ } &
\multicolumn{3}{c|}{\textbf{5-fold Validation} (average)} &
\multicolumn{3}{c}{\textbf{Test Set Evaluation}}
\\\cline{3-8}

& & Precision &
Recall$_{rel}$ & F$_3$ & Precision &
Recall$_{rel}$ & F$_3$\\
& &  (in \%) &  (in \%)& (in \%)
&  (in \%) &  (in \%)& (in \%)
\\
\toprule
0.7 & 0.8 & 79.47  & 40.73   & 72.57  & 78.38  & 36.25   & 70.22  \\
0.6 & 0.7 & 76.07  & 49.75   & 72.25  & 70.93  & 40.25   & 65.90  \\
0.6 & 0.8 & 78.47  & 41.46   & 72.04  & 77.42  & 36.00   & 70.22  \\
0.8 & 0.8 & 78.38  & 41.35   & 71.94  & 82.65  & 40.50   & 74.86  \\
0.5 & 0.8 & 78.15  & 40.08   & 71.37  & 74.29  & 32.50   & 65.82  \\
\midrule
\multicolumn{2}{c|}{Baseline (0, 0)} & 49.23   & 100  & 51.86  & 40.07   & 100  & 42.62  \\
\bottomrule
\end{tabular}
\vspace{-4ex}
\end{table}

Table~\ref{tab:precision} summarizes the precision, relative recall
and $F_3$ statistics for the top-$\mathsf{5}$ threshold configurations for both
the $\mathsf{5}$-fold validation and the test set evaluation as described
in \Cref{sec:datanalysis}.
In the bottom row, we also report the statistics for the
\textit{baseline} configuration used to calculate relative recall.
Among the top-$\mathsf{5}$ configurations, \technique's precision on the test set
ranged from $\mathsf{70.93\%}$ to $\mathsf{82.65\%}$ compared to baseline's $\mathsf{40.07\%}$, their
relative recall ranged from $\mathsf{32.5\%}$ to $\mathsf{40.5\%}$ compared to Baseline's $\mathsf{100\%}$,
and their $F_3$ ranged from $\mathsf{65.82\%}$ to $\mathsf{74.86\%}$ compared to Baseline's $\mathsf{42.62\%}$. 
The best configuration on $\mathsf{5}$-fold validation set uses $\mathsf{threshold}_1 = 0.7$
and $\mathsf{threshold}_2 = 0.8$ and achieves $\mathsf{78.38\%}$ precision, $\mathsf{36.25\%}$ relative
recall, and $\mathsf{70.22\%}$ $F_3$ on the test set evaluation.

Comparing the statistics for $\mathsf{5}$-fold validation and test set evaluation, we observe
high parallels in terms of the precision and relative recall, hinting
towards a degree of \emph{domain-invariance} in the learned patterns as
the train dataset and test dataset were collected more than 6 months apart.
\begin{tcolorbox}[sharp corners, boxsep=0pt,left=5pt,right=5pt,top=5pt,bottom=5pt]
The effectiveness of \technique has a degree of domain-invariance and the
best configuration on $\mathsf{5}$-fold validation achieves $\mathsf{78.38\%}$ precision, $\mathsf{36.25\%}$ relative
recall, and $\mathsf{70.22\%}$ $F_3$ on the test set evaluation.
\end{tcolorbox}

\noindent\textbf{\emph{RQ2: What are the edit sequence patterns learned by \technique?}}
\vspace{0.1cm}

\noindent\textbf{Categorizing ESPs.}
Fixing the best configuration ($\mathsf{threshold}_1 = 0.7$ and  $\mathsf{threshold}_2 = 0.8$)
from the previous study, \technique learned $135$ edit sequence patterns.
We classified these patterns into four categories using the coding
methodology from Section~\ref{sec:RQ2Setup}, as shown in
Table~\ref{tab:categories}. 
\begin{compactitem}
\item \emph{Workflows.}
    We classified $\mathsf{25.9\%}$ of the ESP as \emph{Workflow}, which consists of an
    ESP that describe the workflow of a developer performing a particular
    high-level task.
    For instance, to rename a variable, the developer first renames the variable
    in the declaration, and then renames each one of the variable uses.
    Each step of the workflow is represented as an edit template in the ESP.
    The ESP detects the step that the developer is in the task, and predicts the
    next steps to finish it.
\item \emph{Repeats.} $\mathsf{27.5\%}$ of the ESP were classified as \emph{Repeat},
    which consists of ESP that represent a developer performing a single task
    multiple times.
    For instance, a developer performs an edit to remove the qualifier ``this'' from multiple
    parts of the code.
    The ESP detects that the developer performed the edit in one location and when they move to another similar location, the ESP predicts the change.
\item \emph{Transients and Noise.}
    Finally, we identified two categories of ESP that are not useful:
    \emph{Transient} and \emph{Noise}.
    The former ($\mathsf{13\%}$) relates to edits that are too fine-grained, such as inserting
    \code{publicclass} and then changing to \code{public class}.
    The latter ($\mathsf{33.6\%}$) relates to changes that do not represent a high-level task,
    such as adding a specific switch statement and then adding a break statement.  
\end{compactitem}
After removing noisy ESPs and accounting for ESPs that are variations of a
single type of refactoring, we get $\mathsf{51}$ unique pattern types.
The list of these $\mathsf{51}$ pattern types is shown in supplementary material.

\noindent\textbf{Relating ESPs to IDE features.} 
We further sub-classified the Workflow and Repeat ESPs into two categories:
\emph{Existing Feature} and \emph{New Feature}.
Table~\ref{tab:tasks} presents a list of $\mathsf{20}$ learned ESPs, $10$ existing features
and $10$ new features.
In the first category, we include ESPs that have a corresponding automated tool
or refactoring implemented in the IDE, which allows the IDE to automate the
complete edit in one step using just the spatial context. 
Existing Feature ESPs correspond to $\mathsf{53\%}$ and $\mathsf{39\%}$ of the Workflow and
Repeat ESPs, respectively.
The New Feature category includes ESPs for which we did not find a corresponding
IDE feature.
For instance, ESP $9$ is the inverse of ESP $6$, 
Instead of adding a property and its corresponding parameter,
the developers removes the property and the parameter.
Note that this pattern shows that developers perform changes in a non-standard
way---the ESP first deletes the parameter, then the assignment, and finally
the property.
%

\noindent\textbf{Analyzing Existing Feature ESPs.}
We discuss existing feature ESPs in detail here as they are closely connected
to our motivation of addressing the late-awareness and discoverability problems.
\technique learns existing feature ESPs only because developers manually
performed these edits, which  created a trace of fine-grained edits, instead
of using the IDE tool support, which would lead to a single, larger edit. 
For instance, ESP $4$ represents the edit sequence shown in Figure~\ref{fig:del-param-code}.
The complete edit is automated in one step by Visual Studio (Delete Parameter) using the
spatial context.
To apply this refactoring, the developer needs to put the cursor on the parameter list,
then click on the \emph{Quick Actions} pop-up (the screwdriver icon on the left side of the text pane)
select ``Change signature...'' among all the code edit options,
and then select the parameter to delete.
%
%
To apply this refactoring, not only does the developer need to be aware of this tool,
but also needs to use the tool before making any changes manually.
If the developer starts by deleting the parameter from the parameter list, Visual Studio
will not generate a suggestion to finish the edit sequence by deleting the corresponding
arguments.
Meanwhile, ESP $4$ uses the fact that the developer manually deleted the parameter 
to predict that the developer will delete the corresponding argument--the developer
can use a tool based on ESP $4$ even if they have already started making changes.

%
These results suggest that IDE features were under used, in congruence with
the observation made by \citet{Ge:ICSE:RefactoringSteps}.
They point to the fact that these tools are hard to discover
(discoverability issue)
and even when they are discoverable, developers do not realize the possibility
of using it at the time when that suggestion is available (late-awareness).
\technique can alleviate these problems by producing code edit suggestions using the learned ESPs as shown in Figure~\ref{fig:add-prop-vs}, while developer is editing.



\begin{tcolorbox}[sharp corners, boxsep=0pt,left=5pt,right=5pt,top=5pt,bottom=5pt]
Our qualitative analysis shows that ESPs can be used not only to complete edits when developers typically miss the opportunity to use the IDE tool support but also to predict edits based on new patterns that have no tool support at all.
\end{tcolorbox}

\begin{table}[]
\fontsize{7pt}{8pt}\selectfont
\caption{Categories of Edit Sequence Patterns}
\label{tab:categories}
\vspace{-3ex}
\begin{tabular}{p{0.3cm}p{0.9cm}p{5.3cm}p{5cm}p{0.5cm}}
\textbf{Id} & \textbf{Category} & \textbf{Description} & \textbf{Examples} & $\%$   \\
\toprule
$1$
& Workflow
& A pattern that represents the workflow that a developer performs to complete a task. 
Each edit in the edit sequence represent a step that the developer took. The temporal context is used to detect the previous steps that the developer performed and predict the remaining ones. 
& (i) Developer changes the name of a variable in its declaration and then renames each one of the references (rename variable);

 (ii) Developer changes the type of a variable in the left-hand side of an assignment and then changes the name of the constructor in the right-hand side of the assignment.

& $25.9$ 
\\ 
$2$
& Repeat
& A pattern that represents the developer performing the same task multiple times. All edits in the sequence have the same edit template. The temporal context can predict that the developer will perform the task again. 
& (i) A developer deletes the keyword \texttt{this} from multiple locations in a class (Remove this);

(ii) Developer replaces a static method invocation with a virtual method invocation in multiple locations.

& $27.5$ 
\\
\midrule
$3$
&Transient                  & The sequence represented is too fine-grained to be considered useful.             
& (i) Developer inserts i, then changes to i++;
 
(ii) Developer writes ``publicclass'' then changes to ``public class''
& $13.0$ 
\\
$4$
&Noise                      & We could not identify a high-level task for this pattern.             &  (i) Developer creates a switch statement with a specific case, and then adds a break statement;

(ii) developer cuts and pastes a statement.
& $33.6$
\\
\bottomrule
\end{tabular}
\vspace{-3ex}
\end{table}


\begin{table}[htbp]
\centering
\renewcommand{\arraystretch}{1.5}
\fontsize{7pt}{8pt}\selectfont
\caption{Sample of 20 Edit Sequence Patterns Learned by \technique \label{tab:tasks}}
\vspace{-4ex}
\begin{threeparttable}[t]
\begin{tabular}{llll}
\textbf{Category}              & \textbf{Id}            & \textbf{Pattern Description} & \textbf{Related Feature}\\
\toprule
\multirow{14}{*}{\makecell[l]{Workflow}}
                                      & 1 & Rename method decl $\rightarrow$ Rename method calls  & \textit{Rename Method} \\
                                      & 2 & Insert variable decl $\rightarrow$ Replace constants with new variable & \textit{Introduce Local Variable} \\
                                      & 3 & Insert parameter $\rightarrow$ Insert argument to callsites  & \textit{Insert parameter} \\
                                      & 4 & Delete parameter $\rightarrow$ Delete argument from callsites & \textit{Delete parameter} \\
                                      & 5 & Replace variable declaration by assignment $\rightarrow$ Insert new field  &\textit{Promote local variable to field}\\
                                       & 6 & Insert Property $\rightarrow$ Insert Parameter $\rightarrow$ Insert Assignment &\textit{Initialize property} \\
                                     & 7 & Insert expression $\rightarrow$ Replace it by assignment & \textit{Introduce variable} \\
                                     & 8 & Change type in variable decl $\rightarrow$ Change constructor name in initializer &\textit{New Feature} \\
                                     & 9 & Delete parameter $\rightarrow$ Delete assignment $\rightarrow$ Delete property &\textit{New Feature} \\
                                     & 10 & Insert parameter with default value $\rightarrow$ Replace constants with parameter  &\textit{New Feature}\\
                                     & 11 & Delete field $\rightarrow$ Delete assignment  &\textit{New Feature}\\
                                     & 12 & Insert argument to callsite $\rightarrow$ Remove default parameter value  &\textit{New Feature}\\
                                     & 13 & Insert variable declaration $\rightarrow$ Insert new variable as argument &\textit{New Feature}\\
                                     & 14 & Insert return statement $\rightarrow$ Delete throw "NotImplementedException"  &\textit{New Feature}\\
\midrule
\multirow{6}{*}{\makecell[l]{Repeat}}  
                                    & 15 & Remove `this' in multiple locations & \textit{Remove unnecessary qualifier ``this''} \\ 
                                    & 16 & Remove `cast' in multiple locations & \textit{Remove unnecessary cast}\\
                                    & 17 & Change from qualified name to simple name & \textit{Simplify Name Access} \\ 
                                    & 18 & Converting static method calls to virtual in multiple locations  & \textit{New Feature} \\ 
                                    & 19 & Remove a method invocation from many locations & \textit{New Feature}\\
                                    & 20 & Replace an expression by a method invocation & \textit{New Feature} \\

\bottomrule
\end{tabular}
\end{threeparttable}
\renewcommand{\arraystretch}{1}
\vspace{-3ex}
\end{table}

\ignore{
\begin{figure}
  \footnotesize
\begin{subfigure}[t]{0.24\textwidth}
\begin{lstlisting}[
  language=diff,
  basicstyle=\ttfamily\scriptsize
]
// Edit 1
+ int LargeNum = 1000;                         
  if (IsEmpty(a) && IsEmpty(b)) {
    return 1000;
  } else if (IsEmpty(a)) {
    return 1000 / b.Length;
  }
\end{lstlisting}
\end{subfigure}
\hfill
\begin{subfigure}[t]{0.21\textwidth}
\begin{lstlisting}[
  language=diff,
  basicstyle=\ttfamily\scriptsize
]
// Edit 2
  int LargeNum = 1000;
  if (IsEmpty(a) && IsEmpty(b)) {
-   return 1000;
+   return LargeNum;                           
  } else if (IsEmpty(a)) {
    return 1000 / b.Length;
  }
\end{lstlisting}
\end{subfigure}
\begin{subfigure}[t]{\linewidth}
\begin{lstlisting}[
  language=diff,
  basicstyle=\ttfamily\scriptsize
]
// Edit 3  
  int LargeNum = 1000;
  if (IsEmpty(a) && IsEmpty(b)) {
    return LargeNum;                           
  } else if (IsEmpty(a)) {
-    return 1000 / b.Length;
+    return LargeNum / b.Length; 
  }
\end{lstlisting}
\end{subfigure}
  \caption{Developer manually applies the introduce constant refactoring.}
  \label{fig:extract-constant}
\end{figure}
}

\section{Discussion}
\label{s:discussion}

Our empirical study shows how \technique can help developers by predicting edits based on the temporal context. In this sections, we discuss our results. 

\subsection{\technique's effectiveness compared to related tools}

To our knowledge, there is no other technique or tool that addresses the same problem
so that we can directly compare the precision.
Instead, we indirectly contrast \technique to the closest tool, \texttt{C$^3$PO}~\cite{c3po}.
%
Given a code snippet that is partially edited, \texttt{C$^3$PO} uses a neural model
to predict a completion of the edit within the surrounding $10$ lines of code. 
While not directly comparable to \technique's
70.22\% $F_3$,~\cite{c3po} report C$^3$PO's accuracy 
as 53\%.
C$^3$PO is also restricted to code updates or code deletions, and cannot handle
authoring/adding of new code (e.g., \textit{Insert-Property-Parameter-Assignment}).
%
%
%
%
While edit completion can be seen as an ESP of length $2$, their intended context is
completely different: C$^3$PO is trained on edits from source control, while \technique
is targeted at much more fine-grained edits in the IDE.
%
Further, \technique is driven by temporal context and can predict edits that are spatially
far away from the current editing location (e.g., \textit{Delete parameter \& drop args
at callsites}).

\subsection{Limitations}

Our qualitative analysis revealed some limitations of \technique. First we observed that some ESPs could be represented as a single more general ESP, which happened when two sketches could be generalized into a single sketch. A possible solution for this problem is to try to merge top-ranked ESPs of each sketches in a second round of hierarchical clustering, filtering and selection.

Additionally, we identified several transient and noisy ESPs. While only 13\% of the ESPs were considered transient, and this number can be reduced by implementing a few heuristics based on these patterns, noisy ESPs represented 33.6\% of the patterns. We observed that many of these noisy ESPs were related to patterns that were too specific, which would not produce false positives, but also would not likely trigger on other codebases. This problem can be alleviated with more data and a higher threshold for the support of each edge in the quotient graph (we used support = 2). 

Finally, our ranking step (Section~\ref{sec:rank_esp}) will filter out ESPs with unbounded holes that cannot be filled by hole predicates because these ESPs will have zero precision. For example, our approach will filter out the ESP when user copies a property declaration and then tries to update the name of the copied property ($\version_6 \to \version_7 \to \version_8$ in Figure~\ref{fig:dev-session2}) because we do not know what will be the new name of the property.
Additionally, we rank our ESPs using their performance on the training dataset. However, we could rank them adaptively every time when making a prediction given the context of the code. 
We foresee that combining large language models with our ESPs will be a fruitful research direction because the language models can help fill in unbounded holes in ESPs and rank ESPs adaptively for each prediction.


\subsection{Threats to validity}
\subsubsection*{Construct Validity}
We measured precision by checking if the edit predicted by \technique leads to the exact same code of a subsequent version. Some correct predictions, however, may lead to code that is similar but not exact the same. If the prediction adds a property to the beginning of the class but the developer ended up ending it somewhere in the middle, we will classify the prediction as a false positive, even though both edits are semantically equivalent. Therefore, our false positives may contain some true positives.

\subsubsection*{Internal Validity}
The choice of some parameters used in our evaluation may impact the results. To reduce the bias on the choice of parameters, we performed a cross-validation to select the parameters ($threshold_1$ and $threshold_2$) that would impact the most on the precision of \technique.

%


\subsubsection*{External Validity} \technique learned ESPs from 682 code development sessions from 12 Visual Studio users (developers) that worked on 4 separate C\# code bases. While our results suggest that the ESPs generalize to other datasets, such as our test dataset collected several months after the training dataset, these ESPs might not generalize to other IDEs or programming languages. 
Nevertheless, we identified several ESPs that have corresponding features in the IDE, and the proposed technique is independent of programming languages. 

\subsubsection*{Verifiability} Our dataset is not publicly available due to a non-disclosure agreement with our participants.

\ignore{

\paragraph{Multiple Refactoring}
These are a series of refactoring that developers perform, \todo{where the edits may or may not have a common parameter}.
For instance, in a deeply nested call chain, often a single parameter is used at the bottom of the call chain is passed through every method above the call (see
  Figure~\ref{fig:delete-param-casacade}).
  This style is sometimes considered an anti-pattern and developers often
  refactor code to avoid this style.
  \begin{figure}[htbp]
  \includegraphics[scale=0.53]{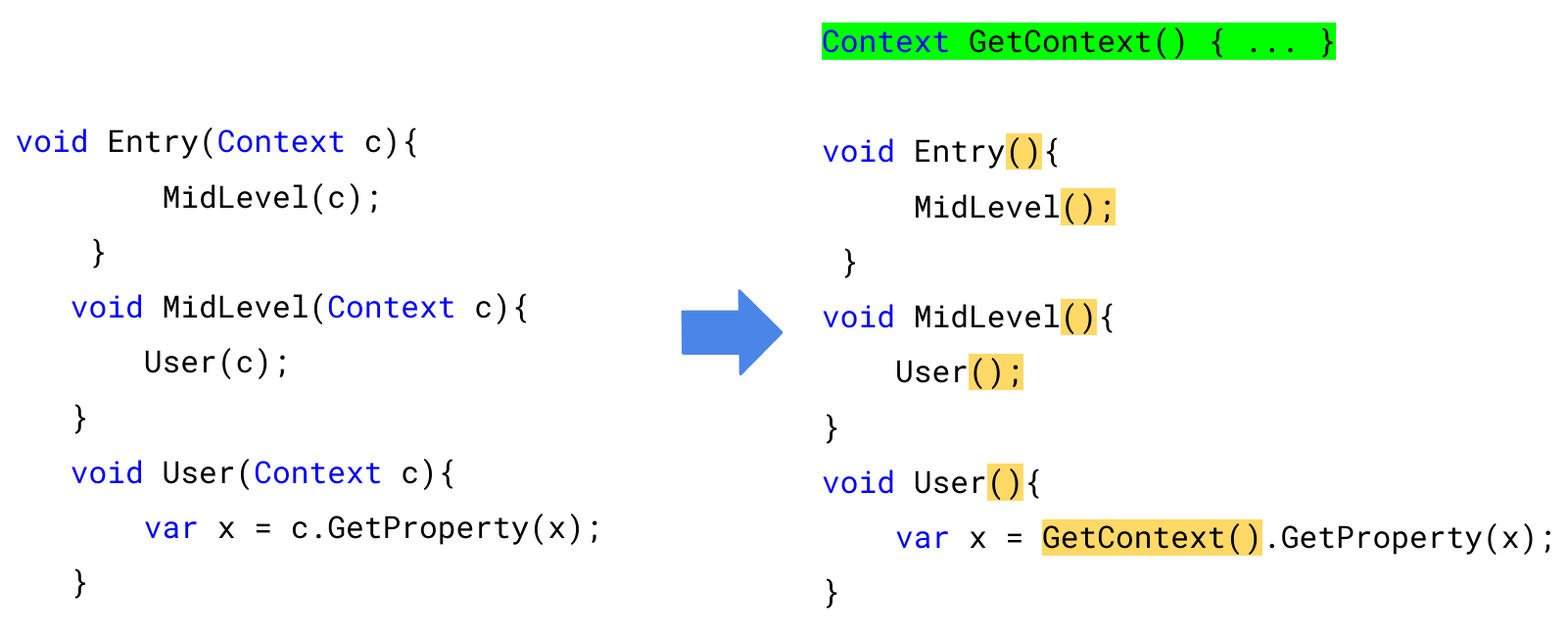}
  \vspace{-1mm}
  \caption{Change Method Return type}\label{fig:delete-param-casacade}
  \vspace{-2mm}
\end{figure}
  A common workflow to achieve this:
  \begin{inparaenum}[(a)] 
    \item Delete parameter from entry point (here \code{Entry});
    \item Delete its usages (here delete arguments at the call to \code{MidLevel1}); 
    \item Propagate this change by deleting the parameter of the corresponding method declaration (here definition of \code{MidLevel1});
    \item Repeat until the bottom of the call chain is reached.
  \end{inparaenum}
While most current IDEs support offer the \textit{Change Method Signature} feature to add or remove method parameter, they are inept to assist a developer to perform such chained refactorings.
Similar to the observation made by the previous researchers~\cite{Ketkar:UnderstandingTypeChanges}, we too observed that a type change refactoring leads to another type change refactoring (to propagate type constraints) and rename refactoring (to maintain consistency).
We also observed tasks, where the series of refactoring performed is not inter-dependent upon each other.
For instance, the developers converted all the explicit types in a method block to use \code{var} (e.g., \smcode{string a ="A";}$\rightarrow$\smcode{var a ="A";}) or apply a naming convention to methods and fields in a class.



\paragraph{Multiple- Quick action}
We observed that developers often perform a code-fix (or quick action) multiple times at different places. For instance, the developers converted all the explicit types in a method block to use \code{var} (e.g., \smcode{string a ="A";}$\rightarrow$\smcode{var a ="A";}).

\paragraph{Code Deletion Tasks}

\paragraph{IDE Intelligence}

We found 12 ESPs where the developer performs an edit based on a recently performed edit.
We further grouped these 12 patterns into 9 distinct tasks like \textit{Create a variable and iterate over it} or \textit{Create a variable and then invoke a instance method}. 
Visual Studio assist developers to perform these tasks (partially) by offering AI-powered feature (like \textit{IntelliSense}), to provide advanced code-completion for invoking an instance method. 
We observed that, often developers first use a symbol (identifier) and then declare it as a method or variable.
While IDEs assist at such a workflow with the ``Generate Code From Usage'' feature, we found several instances where the developer creates the declaration by hand.    
We also found repetitive ESPs, where the developers perform the same edits at multiple places like systematically updating the usage of an API. 
While these repetitive patterns are automated by \textit{IntelliCode} (\Cref{fig:bluepencil}), we found scenarios where \textit{IntelliCode} does not currently assist, but could benefit from. For instance, updating multiple  \textit{string} or \textit{number} literals similarly. 


\subsection{Implications}
We discuss some main insights and learnings we gained from analyzing code edit data
using \technique.

\smallskip\noindent{\textbf{Existing automation tools are not used at their full potential.}}
\\
{\em{Observation:}} Modern IDEs provide support for automating several common code edits, refactorings,
or code fixes, but we still observe developers doing those changes by hand. 
\\
{\em{Analysis:}} We believe the reason is that this automation is hard to discover 
(discoverability issue), and even when it is discoverable, developer do not realize 
the possibility of using it at the time when that suggestion is available (late-awareness). 
\\
{\em{Action:}} The ESPs for tasks that are currently automated in IDEs
can be monitored and we can build telemetry to check how often people are not using available
tools.

\smallskip\noindent{\textbf{Improving current automation.}}
\\
{\em{Observation:}} We found new ESPs, along with some well-known patterns that are
currently automated.
\\
{\em{Analysis:}} There are a few automation tools ([BeneFactor, WitchDoctor, IntelliCode suggestions] 
that exploit temporal context. They observe the past edit(s) made by the developer to surface
the next edit suggestion. This way they are able to address discoverability and late-awareness 
issues. 
\\
{\em{Action:}} The new spatiotemporal ESPs we found can be used in the same way to
make edit suggestions.  The suggested edits can just be those that IDEs already implement, such
as quick actions in Visual Studio. They suggested edits can also be new quick actions. The main
benefit would be improved discoverability and eliminating late-awareness issues.

\smallskip\noindent{\textbf{Novel ways to exploit spatiotemporal context.}}
\\
{\em{Observation:}} Analysis of the edit sequences performed by developers revealed some patterns that do not directly fall into any class of existing automated code fixes or refactorings.
\\
{\em{Analysis:}} The new patterns suggest new ways to use the spatiotemporal context to suggestion edits.  
\\
{\em{Action:}} First, it may be possible to suggest a {\em{composition of quick actions}}; for example, 
if a developer has added 4 parameters to a constructor, we can make {\em{one}} suggested edit that
handles all 4 parameters simultaneously (as opposed to having the developer invoke some quick action 
4 times). The main benefit would be that we get automation of larger workflows without needing to implement a larger quick action.
\\
{\em{Action:}} Second, temporal context could be used to rank or filter out quick actions.
For example, when a developer changes the type of a returned expression, then
instead of showing the action to fix the expression, we show the action to fix the return type 
of the method.
As another example, when a develper creates a variable, we can filter out the
complaint that the variable is unused because in all likelihood the developer would
next use the variable. The main benefit here is that we are able to reduce noise 
in the UI.
\\
{\em{Action:}} Finally, copy-paste is a common pattern used for authoring code,
and temporal context can yield a smart copy and paste experience.
For example, after a developer does a copy-and-paste on a property, they next update 
the type and name in the pasted code.  
After a developer does a copy-and-paste on an if-condition, they update the variables next.
The temporal context can thus be used to help the developer modify the pasted code.
The main benefit is that we can help developers author code when copying and pasting. 

\smallskip\noindent{\textbf{Data-driven handling of feature requests.}}
\\\ameya{Remove?}
{\em{Observation:}} Several open feature requests for automation in IDEs can be mapped to
spatiotemporal ESPs we extracted using \technique.
\\
{\em{Analysis:}} The ESPs generalize the example provided in the feature request and also provide details on their frequency.
\\
{\em{Action:}} Use information from the ESPs to prioritize feature requests, and then use the temporal context to suggest the generalized edits in the pattern.
}

\ignore{

\paragraph{Insert Code Snippet}
We found \noOfInsertSnippetsPatterns temporal edit patterns where the developer inserts new code snippets (like statements methods, or properties), that we unified to   \noOfInsertSnippets  \textit{temporal edit patterns}.
One pattern we observed was - \\ 
$\mathsf{AddProp}(\mathsf{type}, \mathsf{propName})
\cdot \mathsf{AddGet}(\mathsf{propName})
\cdot \mathsf{AddSet}(\mathsf{propName}))$ or $ \mathsf{AddProp}(\mathsf{type}, \mathsf{propName})
\cdot \mathsf{Update}(\mathsf{propName, newPropName})
\cdot \mathsf{Update}(\mathsf{type}, \mathsf{otherType}))$,

This \textit{temporal edit pattern} expresses two ways to insert a new property - (i) token by token by hand (ii) \textit{Copy-Paste-Adapt} - duplicate an existing property and then update this duplicated property.
For instance, in \Cref{fig:copy-paste-update} the developer first duplicated the property \code{int X}, and then updated the duplicate property's name to \code{Y}.
Modern IDEs provide the \textit{code snippet} feature to insert ready-made snippets of code you can quickly into the code, that can potentially assists at most tasks in this category.
We did not observe any usage of this feature create new code snippets, rather we observed that developers apply \textit{copy-paste-adapt} to create new code. 
\begin{figure}[htbp]
  \includegraphics[scale=0.55]{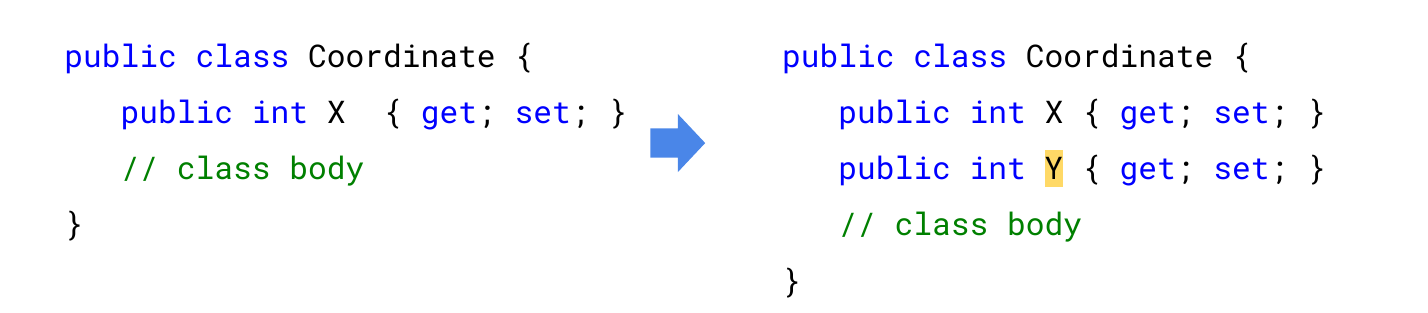}
  \vspace{-1mm}
  \caption{Copy-paste and repetitive update}\label{fig:copy-paste-update}
  \vspace{-2mm}
\end{figure}
We observed that the developers often employ this workflow to create many syntactically similar code snippets - like similar properties in a class or similar \textit{assert} statements in a test case.
%


\paragraph{Refactoring} These tasks are the well known refactoring that developers performed by hand.
For instance we observed the a \textit{temporal edit pattern} to apply \textit{Extract Constant} refactoring - 
$\mathsf{AddVarDecl}(\mathsf{type}, \mathsf{varName}, constVal)
\cdot \mathsf{Update}(\mathsf{constVal, varName})*$ 
In \Cref{fig:extract_constant}, instead of invoking the available \textit{Extract Variable} refactoring, the developer first inserts a new variable declaration and performs two consecutive edits to update the literal \smcode{1000} to \smcode{Limit}.
While IDEs automate many refactorings like \textit{Insert/Delete Method parameter} or \textit{Move method}, we found that developer perform them by hand.
Our qualitative analysis result is in congruence with the observation made by researchers Ge et al.~\cite{Ge:ICSE:RefactoringSteps}, that refactoring tools are under used. 
This highlights new opportunities for IDE designers to invest into developing tools that detect the developers intention to perform a refactoring and offer to complete the change (like \Cref{fig:manual-rename}).

\begin{figure}[htbp]
  \includegraphics[scale=0.6]{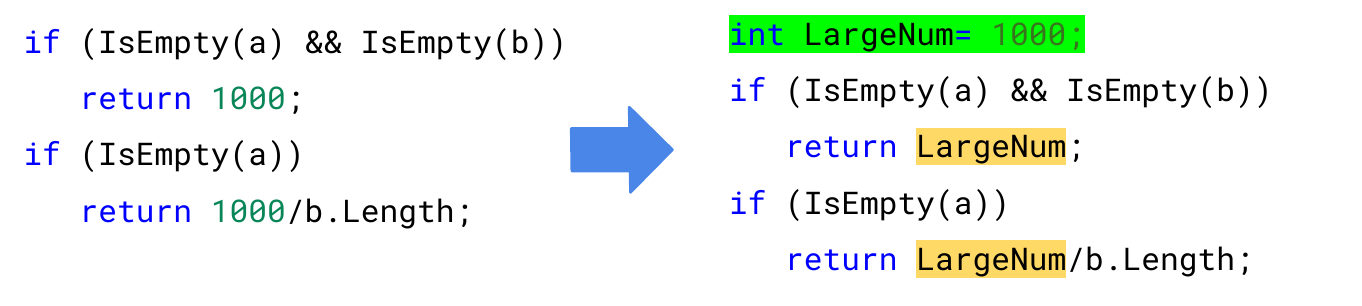}
  \vspace{-1mm}
  \caption{Change Method Return type}\label{fig:extract_constant}
  \vspace{-2mm}
\end{figure}

}
\section{RELATED WORK}
Our work distinguishes itself from existing work by simultaneously 
\textbf{learning} (i) the edits from code development sessions in IDEs, and (ii) the temporal relation between the edits. 
Existing work on {\em{edit patterns}} is mostly focused on coarse-grained edits,
whereas existing work on {\em{temporal patterns}} is limited in the scope of edits it considers.


\noindent \textbf{Learning Edit Patterns from Commits}
Previous researchers have proposed a plethora of techniques that learn edits patterns from the coarse commit level changes. 
\citet{revisar} proposed a technique that infers code change pattern as rewrite rules (not specific fixes, or bugs) using anti-unification and a greedy algorithm for clustering. 
Similarly, \citet{getafix} proposed a technique (\texttt{Getafix}) to learn bug fix patterns using anti-unification. They presented a novel hierarchical, agglomerative clustering algorithm to cluster the examples. Getafix then applies an effective ranking technique that uses three metrics to produce a small and appropriate number of fixes for a given bug.
\texttt{Getafix} inspires our design of the algorithms synthesizing edit sequence patterns, including anti-unification and the hierarchical clustering algorithm. However, \technique learns edit sequence patterns (not just an edit template) from fine-grained code development sessions instead of commit level data.

Recently, \citet{ketkar2022tcinfer} proposed \texttt{TCI-Infer} that learns the rewrite rules to perform type changes from the type changes identified by \texttt{RefactoringMiner}\cite{Tsantalis:TSE:2020:RefactoringMiner2.0} in the commit level history of Java projects. 
Similarly, \texttt{A3}~\cite{A3:lamothe:tse} and \texttt{MEditor}~\cite{Meditor:ICPC:2019} infer the adaptations required to perform library migration by analyzing the changed control/data flow across the commit.
\citet{Kim:SystematicCodechange} proposed a syntactic approach to automatically discover and represent systematic changes as logic rules with the goal to enhance developer's understanding about the program's evolution.
Previously, \citet{REPERTOIRE:Ray, Mengsydit:creating, Meng:LASE} proposed techniques to perform systematic code changes by creating a context-aware edit script, finding potential locations and transforming the code.
\citet{genericPatch} propose a technique that generates generic patches from a set of files and their updated versions and applies these to other files.
\citet{yin2019learning} propose a model that combines neural encoder with edit encoder, to express salient information of an edit and can be used to apply the edit. 
In contrast, \technique learns \textit{sequences} of edit patterns (as opposed to a single edit) that developers often apply while performing their daily code development activities in an IDE.
Blue-Pencil~\cite{Miltner:BluePencil} identifies repetitive changes, and automatically suggests similar repetitive edits. 
However, all these techniques either focus on a specific kind of edit or operate on coarse-grained VCS data which is \textit{imprecise}, \textit{incomplete} and makes it \textit{impossible} to involve the temporal aspect~\cite{PerilsVCS:Negara}.

\noindent  \textbf{Learning Edit Patterns from Sequence of Changes.}
\citet{Mesbah:deepDelta} propose \texttt{DeepDelta} to automatically suggest code fix for the common classes of build-time compilation failures.
They encode the human-authored, in-progress changes into a domain-specific language and feed them to a neural machine translation network along with the compiler diagnostic. 
In contrast, \technique operates over finer IDE level sequences of code changes to learn the sequences of edits performed for a large variety of daily code development activities, not limited to compilation error fixes.
Previously \citet{Negara:ICSE:MiningEdits} proposed a technique to detect \textit{high-level} code change patterns from the \textit{fine-grained} sequences of edits recorded in the IDE.
In contrast, \technique learns sequences of \textit{executable} \textit{edit templates} that not only captures the \textit{high-level} code change patterns but also the different workflows or sequences of edits that were applied to perform the change. 

\noindent  \textbf{Detecting Sequences of Edit Patterns.}
Previous researchers have also tackled the temporal aspect of applying edits like \technique. \cite{Ge:ICSE:RefactoringSteps,witchdoctor} identified that \textit{discoverability} and \textit{late-awareness} led to \textit{underuse} of refactoring tools, and proposed the techniques \texttt{Benefactor} and \texttt{WitchDoctor} to overcome it.
The users of these technique have to manually encode the sequence of edits that are applied to perform a larger refactoring. The authors conducted interviews with software developers to manually recorded the sequences of edits they apply to perform a refactoring. These tools detect when the user is performing a refactoring, and suggest a completion.
On the other hand \technique, automatically learns the sequence of edits that developers apply to perform any high level programming task (beyond refactorings) from code development sessions. \technique basically automates the entire pipeline proposed by \texttt{Benefactor} and \texttt{WitchDoctor}.

\section{Conclusion}


We introduce and tackle the problem of learning edit sequence patterns, with 
the aim of adding temporal context into IDE edit suggestion tools.
ESPs capture fine-grained details in developers' editing
behaviour, and can be used to address the two key challenges towards
broader usage of IDE edit suggestion tools---discoverability and late awareness.
Our experiments show that \technique can not only learn and automate editing
patterns related to existing IDE tools, but also discover new patterns!
Besides being useful to automatically make edit suggestions in the IDE,
we foresee the ESPs being used by IDE toolsmiths to decide and prioritize
what new features they should develop in a data-driven manner.

ESPs and temporal context can help develop tools that are
more accurate, and in turn, allow for more aggressive presentation of suggestions
with hand-raising interfaces like ``grey text''.
We are currently exploring the design space for these interfaces to determine the
best way to present different kinds of insert, delete, and update edit suggestions
based on the IDE's confidence in those suggestions.
With the advent of powerful pre-trained language models for source code has also
opened up the possibility of extending \technique to cover more general patterns 
with unbounded holes.
Overall, ESPs offer the possibility of developing a new generation of IDE tools
and exploring a rich and exciting set of ideas from a range of research fields
from HCI to AI to static analysis.

\ignore{
In this paper, we developed a technique \technique to capture temporal context in code edits by learning edit sequence patterns from development sessions.
Our evaluation of \technique on test set shows that \technique is effective at predicting edit sequences performed in the IDE with 78\% precision. 
The qualitative analysis shows that 53.4\% of the edit sequence patterns learned by \technique are useful (25.9\% and 27.5\% of them related to \emph{Workflow} and \emph{Repeat} pattern, respectively). Most of the useful patterns are either not supported by Visual Studio or only have a one-step version of the edit sequence pattern. 
The result highlighted opportunities for improving the experience of developers by incorporating the learned edit sequence patterns into IDEs.
Our edit sequence patterns indicate that IDEs can exploit the spatial and temporal cues provided by the developers about
what they will be doing next -- using
the discovered edit sequence patterns -- to make (and hide) suggestions from the developers in a smooth way.
}
\ignore{
The findings of our study depend on the accuracy of \technique to mine temporal edit patterns from the collected code development sessions. 
However, \technique has three main limitations:
\begin{enumerate}
    \item  It does not infer the parameter flow constraints in the temporal edit patterns.
    For instance, in \textit{Create Variable and invoke instance method} (Id: 6) - the developer adds a new local variable and then invokes an instance method upon this variable; \technique cannot guarantee this.
    \item  It reports several duplicate temporal edit patterns. 
    For instance, for the motivating example \technique will return multiple temporal edit patterns, where the first edit is abstracted into different levels of abstraction - "add a public get-set property" or "add a public property".
    \item it uses a very simple embedding to represent the edit, that might not not suffice to express large and complex edits like refactorings.
\end{enumerate}
To mitigate this, the authors manually analyzed the temporal edit sequence patterns and (i) ~annotated them with the parameter flow constrains, (ii)~identified duplicates, and (iii)~ merged the temporal edit sequence patterns that were related to the same edit task but were not merged by \technique.
}




\balance
\bibliography{bibliography}

\end{document}